\documentclass[twocolumn,english]{article}
\usepackage[T1]{fontenc}
\usepackage{amssymb}
\usepackage{graphicx}
\usepackage[switch,columnwise]{lineno}

\makeatletter
\usepackage{a4}
\input{epsfig.sty}
\def\fnum@table{\tablename~{\bf\thetable}}
\def\fnum@figure{\figurename~{\bf\thefigure}}
\def\tablename{\footnotesize{\bf Table}}
\def\figurename{\footnotesize{\bf Figure}}

\def\be{\begin{equation}}
\def\ee{\end{equation}}

\makeatother

\usepackage{babel}
\begin{document}

\title{\textbf{Addressing uncertainties of model predictions for extensive air showers
initiated by high energy cosmic rays}}

\author{Sergey Ostapchenko$^{1}$,  Tanguy Pierog$^{2}$
  and G\"unter Sigl$^{1}$\\
$^1$\textit{\small Universit\"at Hamburg, II Institut f\"ur Theoretische
Physik, 22761 Hamburg, Germany}\\
$^2$\textit{\small  Institute for Astroparticle Physics, Karlsruher Institut
 f\"ur Technologie,}\\
\textit{\small Hermann-von-Helmholtz-Platz 1, 76344 Eggenstein-Leopoldshafen, Germany}\\
}

\maketitle
\begin{center}
\textbf{Abstract}
\par\end{center}
A new Monte Carlo generator of hadronic collisions, QGSb, is applied for studying
model uncertainties regarding calculations of the development of extensive air showers (EAS) initiated by interactions of high energy cosmic rays in the atmosphere.
More specifically, we investigate possibilities to modify the model predictions for two  EAS characteristics mostly used in experimental studies of cosmic ray composition: air shower maximum depth  and the muon number at ground level.  For all the considered modifications of the model, we discuss in some detail the underlying physics mechanisms and investigate the impact of the changes, regarding a comparison of the model results to relevant accelerator data.

\section{Introduction\label{intro.sec}}
Experimental studies of ultra-high energy cosmic rays (UHECRs) are performed by
indirect methods: reconstructing properties of primary
cosmic ray (CR) particles from measured characteristics of
  extensive air showers (EAS) -- nuclear-electromagnetic cascades initiated by
  interactions of primary CRs in the atmosphere \cite{nag00,blu09}. Therefore, an
  accurate  description of the development of such cascades by the corresponding
   numerical tools, like the CORSIKA program \cite{hec98}, is of significant  importance. In particular, studies of mass composition of UHECRs rely substantially on the modeling of interactions with air nuclei of both primary CRs and of secondary hadrons produced in the subsequent cascades of nuclear interactions in the atmosphere, which involves Monte Carlo (MC) generators of hadronic interactions \cite{eng11}.
   
 Consequently, of significant interest is to quantify uncertainties of predictions of such MC generators, regarding EAS development, which was the subject of numerous studies (e.g., \cite{ulr11,par11,ost24c,ost24d}). 
 Generally, one may get some feeling about such uncertainties by analyzing the spread of the corresponding results of current CR interaction models. 
 Yet, because of potential deficiencies of some approaches, the actual uncertainties may appear to be smaller (e.g., \cite{ost16,ost23}). 
 On the contrary, in case some important physics is not accounted for, all current  models may deviate from the true picture.
 
 In \cite{ops25,ops26}, we presented a new MC generator of CR interactions, QGSb, which is based on a relatively simple and transparent theoretical formalism, while being characterized by a sufficient parameter freedom. Thanks to its substantial flexibility, this generator can be used to study uncertainties for EAS predictions, within the limits allowed by relevant accelerator data regarding hadron-proton and hadron-nucleus interactions. In the current work, we exemplify such a study, addressing possibilities to vary the predictions for two EAS characteristics mainly employed in UHECR composition studies: the average maximum depth of air shower development, $X_{\max}$, and the EAS muon number at ground level, $N_{\mu}$. In particular, we identify key model parameters relevant to such studies and discuss in some detail the corresponding interaction mechanisms, addressing also the constraints imposed by relevant accelerator data, as well as the uncertainties of the latter.

\section{Alternative tuning of model parameters  \label{tunes.sec}}

\subsection{Modifying the predicted EAS maximum depth  \label{xmax.sec}}
Let us first investigate possibilities to modify the calculated longitudinal development of extensive
air showers. The knowledge of the corresponding model uncertainties, notably regarding
 predictions for EAS maximum depth $X_{\max}$, are of considerable importance for analyzing 
 and interpreting experimental data from giant air shower arrays.  
 Moreover, recent analysis by the Pierre Auger collaboration indicates that substantially
 larger  $X_{\max}$ values, compared to current model predictions, are required for a 
 self-consistent interpretation of the data, in terms of mass composition of primary
  cosmic rays \cite{abd24}.

\subsubsection{Aiming at a larger $X_{\max}$  \label{xmax-high.sec}}
\begin{figure}[htb]
\centering
\includegraphics[height=6cm,width=0.48\textwidth]{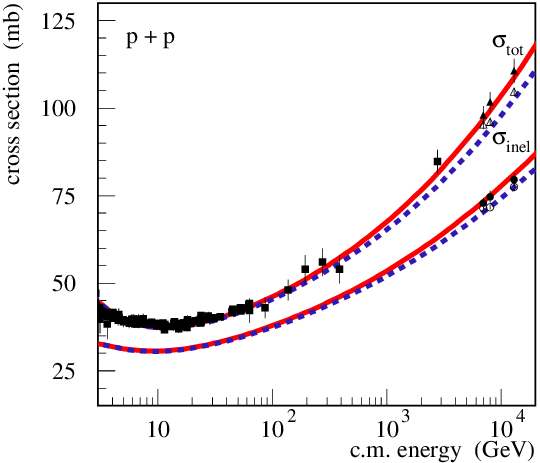}
\caption{Center-of-mass (c.m.) energy dependence of  the total and inelastic
 $pp$   cross sections, as indicated in the plot,
  compared to experimental data \cite{ant19,aad23,pdg} (points).
  The calculations performed using the default model (tune 1)
   or using a smaller value for the  overcriticality
of the ``semihard Pomeron''  (tune 2) are shown, respectively, by red solid 
 and blue dashed lines. The results of the TOTEM  experiment for
  $\sigma^{\rm tot}_{pp}$ and $\sigma^{\rm inel}_{pp}$ are plotted as filled triangles and filled circles, respectively, while the ones of the   ATLAS
 experiment are shown by the corresponding open symbols.}
 \label{fig:sigpp}       
\end{figure}

\begin{figure}[htb]
\centering
\includegraphics[height=6cm,width=0.48\textwidth]{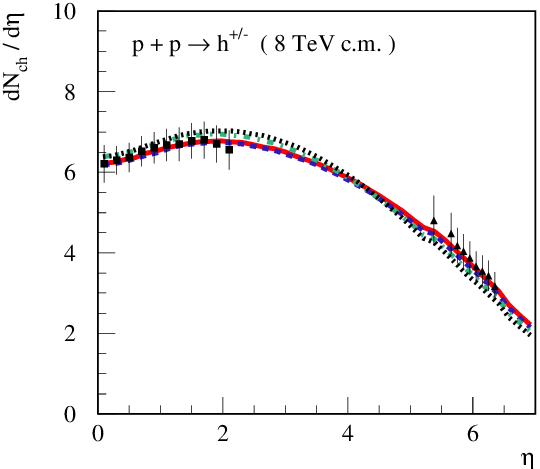}
\caption{Pseudorapidity distribution of charged hadrons in c.m.\ frame, produced in $pp$
collisions at $\sqrt{s}=8$ TeV, as calculated using   tunes 1, 2, 3, and 4 of QGSb --
red solid, blue dashed, green dash-dotted, and black dotted lines, respectively,
  compared to the data of CMS (squares) and TOTEM (triangles) \cite{cha14}.}
 \label{fig:ppeta}       
\end{figure}

Obviously, the simplest way to change  $X_{\max}$ predictions is to modify the energy dependence
of the inelastic proton-air cross section, $\sigma^{\rm inel}_{p-{\rm air}}$, which controls
the mean free path of the primary CR proton in the atmosphere (e.g., \cite{ulr11}).
 However, $\sigma^{\rm inel}_{p-{\rm air}}$
 is rather strongly constrained by measurements of  proton-proton interaction
cross sections at the Large Hadron Collider (LHC) since the quantities are closely related
to each other by the very successful Glauber-Gribov formalism \cite{gla56,gri69}.
While the default settings of the QGSb model, hereafter referred to as tune~1,
 make it  consistent with the results of the TOTEM
experiment \cite{ant19} on the total  and inelastic $pp$ cross sections, as one can see
in Fig.\ \ref{fig:sigpp},
 one can alternatively tune the relevant model parameters, 
 using measurements of the ATLAS experiment \cite{aad23},
shown by open symbols in  Fig.\ \ref{fig:sigpp},
which indicated a slightly slower energy rise of $\sigma^{\rm tot/inel}_{pp}$.
To this end, we choose a $\simeq 5$\% lower  overcriticality
of the  ``semihard Pomeron'',  $\Delta_{\rm sh}=0.21$, which governs the energy 
dependence of the calculated cross sections in the very high energy limit \cite{ops26}.
The corresponding results for   the total and inelastic $pp$ 
 cross sections are shown in Fig.\ \ref{fig:sigpp} as dashed lines.
  It is noteworthy that such  a change of the parameter  $\Delta_{\rm sh}$
reduces also the predicted multiple scattering rate  in proton-proton and proton-nucleus
collisions (see Eqs.\ (3) and  (6)   in \cite{ops26}). Therefore, to stay in agreement with LHC
data on charged hadron production in the central rapidity range, we have to make changes 
in the hadronization procedure of the model: by increasing the density of produced hadrons
per unit rapidity. This is achieved by choosing a $\simeq 30$\% larger value of the string 
fragmentation parameter for semihard Pomerons, $\Lambda ^{\rm (sh)}=4.6$
 (cf.\ Eq.\ (36) and Table 6 in \cite{ops26}). The pseudorapidity $\eta$
 distribution of charged hadrons, $dN^{\rm ch}_{pp}/d\eta$,
  produced in $pp$ collisions at $\sqrt{s}=8$ TeV,
 calculated both with  the default model or using the above-discussed modifications
 of the model parameters (hereafter referred to as tune 2), are plotted in Fig.\  \ref{fig:ppeta}, 
 in comparison to the respective data of the CMS and TOTEM experiments.

\begin{figure}[htb]
\centering
\includegraphics[height=6cm,width=0.48\textwidth]{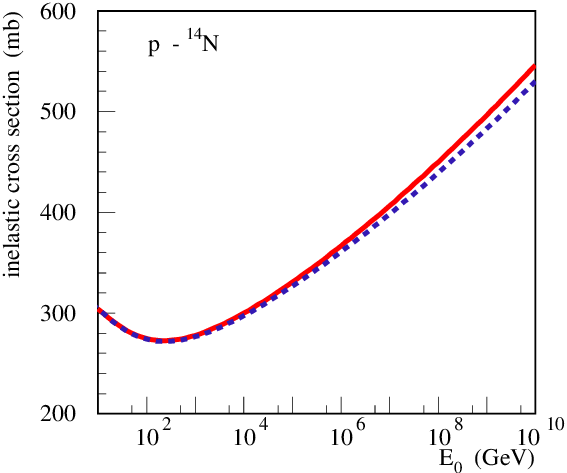}
\caption{Laboratory (lab.) energy dependence of the inelastic cross section for proton-nitrogen
 collisions. The notations for the  lines are the same as in  Fig.\ \ref{fig:sigpp}.}
 \label{fig:siga}       
\end{figure}

\begin{figure}[htb]
\centering
\includegraphics[height=6cm,width=0.48\textwidth]{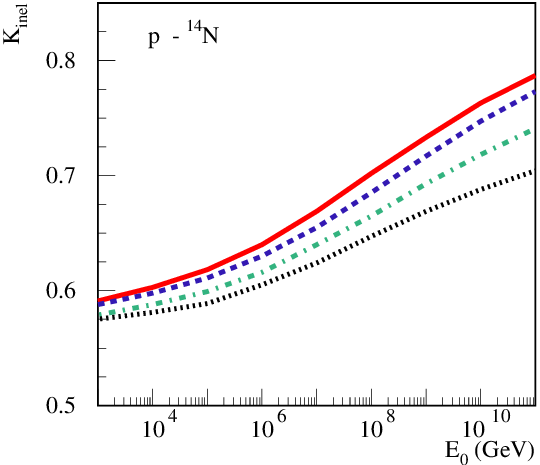}
\caption{Lab.\ energy dependence of the inelasticity $K^{\rm inel}_{p{\rm N}}$
of  proton-nitrogen collisions.
 The notations for the  lines are the same as in  Fig.\ \ref{fig:ppeta}.}
 \label{fig:kinel}       
\end{figure}

\begin{figure}[htb]
\centering
\includegraphics[height=6cm,width=0.48\textwidth]{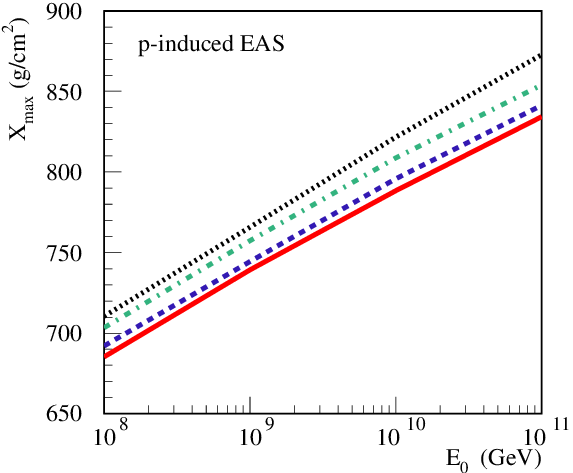}
\caption{Dependence of the average shower maximum depth  $X_{\max}$, 
for $p$-induced vertical EAS, on the primary proton energy $E_0$.
 The notations for the  lines are the same as in  Fig.\ \ref{fig:ppeta}.}
 \label{fig:xmax}       
\end{figure}

Regarding  $\sigma^{\rm inel}_{p-{\rm air}}$, it is much more weakly influenced by the
considered changes, compared to  $\sigma^{\rm inel}_{pp}$, since proton-nucleus
cross sections are largely dominated by the nuclear size, for sufficiently heavy nuclei.
This is illustrated in Fig.\ \ref{fig:siga},
 where we observe up to $\simeq 3$\%
 difference for the inelastic proton-nitrogen interaction cross section 
 $\sigma^{\rm inel}_{p{\rm N}}$ between the tunes 1 and 2 of the model.
Such a modification of the proton-air cross section alone would enlarge average 
 $X_{\max}$ by less than 2 g/cm$^2$. Yet, as discussed above, using a smaller
   overcriticality of the  semihard Pomeron  reduces also the 
   predicted multiple scattering rate  in proton-proton and proton-nucleus
collisions: a smaller number of strings of color field would be attached to the
incident proton. Hence, a smaller fraction of the proton energy would be taken
by string end partons, thereby reducing the inelasticity $K^{\rm inel}$
 (the relative energy loss of ``leading'', i.e., most energetic,
  secondary nucleons)  of $pp$ and $pA$ interactions.
Indeed, for this model tune,    the inelasticity $K^{\rm inel}_{p{\rm N}}$
of  proton-nitrogen collisions is reduced by up to $\simeq 2$\% at the highest energies, compared to the default QGSb (cf.\ solid and dashed lines  in  Fig.\  \ref{fig:kinel}).
Both effects combined give rise to a noticeably larger  average   $X_{\max}$ for
proton-induced EAS: by up to $\simeq 7$ g/cm$^2$ at $E_0=10^{20}$ eV, as one can see 
in   Fig.\  \ref{fig:xmax}.

A more radical reduction of the  inelasticity of proton-air collisions can be
achieved considering softer  momentum distributions of constituent partons 
to which strings of color field are attached \cite{par11,ost24c,ost16,ost03}.
To this end, in addition to using the smaller overcriticality of the  
semihard Pomeron,    $\Delta_{\rm sh}=0.21$, we consider larger values
of the exponent  $\alpha _{\rm sea}^{\rm (sh)}$ describing the momentum distribution
of constituent sea (anti)quarks, $\propto x^{-\alpha _{\rm sea}^{\rm (sh)}} $
(see Eq.\ (30)  in \cite{ops26}), to which 
semihard Pomeron strings are attached: using  
$\alpha _{\rm sea} ^{\rm (sh)}=0.8$ and 0.95, instead of the default
setting   $\alpha _{\rm sea} ^{\rm (sh)}=\alpha _{\mathbb{R}}=0.5$. Here again, 
 to keep the agreement with LHC
data on charged hadron production in the central rapidity range, 
we need to enlarge further the  string 
fragmentation parameter for semihard Pomerons,
using, respectively, $\Lambda ^{\rm (sh)}=6.5$ and 9.5 for the two choices 
of $\alpha _{\rm sea} ^{\rm (sh)}$, which we refer hereafter as tunes 3 and 4.
It is noteworthy that all the considered changes do not affect the model 
agreement with accelerator data corresponding to hadronic interactions
at fixed target energies since those are dominated by soft Pomeron exchanges
and by hadronization of  soft Pomeron strings.
A comparison of the calculated  $dN^{\rm ch}_{pp}/d\eta$ at $\sqrt{s}=8$ TeV 
with the data of 
CMS and TOTEM is shown in  Fig.\  \ref{fig:ppeta}, while the obtained energy
dependence of the  inelasticity of proton-nitrogen collisions is plotted 
 in  Fig.\  \ref{fig:kinel}. As one can see in  Fig.\  \ref{fig:kinel},
  $K^{\rm inel}_{p{\rm N}}$ is reduced by up to 6 and 11\%, respectively,
  for the tunes 3 and 4.  In turn, this allowed us to increase   $X_{\max}$ for
proton-induced EAS by up to $\simeq 20$ and $\simeq 40$  g/cm$^2$, for the two cases,
correspondingly, as shown  in   Fig.\  \ref{fig:xmax}.
 
\begin{figure*}[t]
\centering
\includegraphics[height=11.cm,width=\textwidth]{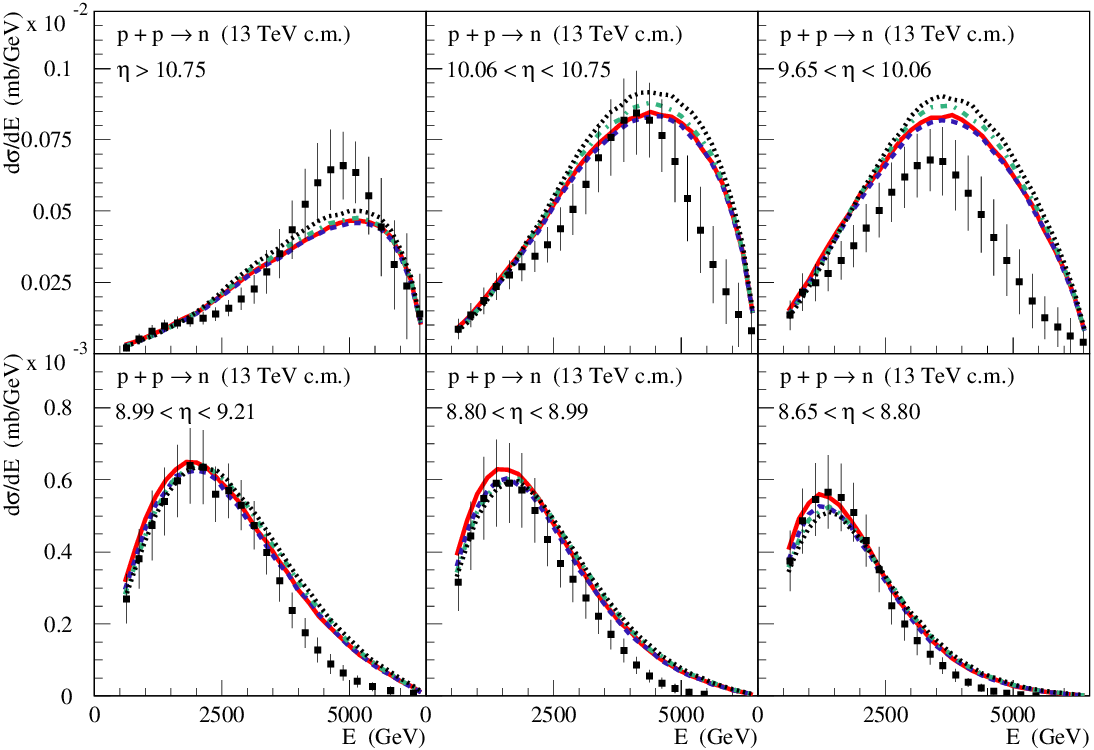}
\caption{Neutron energy spectra in c.m.\ frame,
for  $pp$ collisions at $\sqrt{s}=13$ TeV,
for different pseudorapidity intervals, as indicated
in the plots, in comparison to LHCf data \cite{adr18} (points).
 The notations for the  lines are the same as in  Fig.\ \ref{fig:ppeta}.}
\label{Flo:lhcf-n}       
\end{figure*}

 Regarding a possibility to discriminate between the considered modifications
 of the model,   promising are measurements of forward neutron spectra
 by the LHCf experiment \cite{adr15,adr18}. As one can see in 
 Fig.\  \ref{Flo:lhcf-n}, the trend towards larger $\alpha _{\rm sea} ^{\rm (sh)}$
is somewhat disfavored  by the data: the predicted forward neutron yield  exceeds
 the measured one overall. 

\begin{figure}[htb]
\centering
\includegraphics[height=6.cm,width=0.48\textwidth]{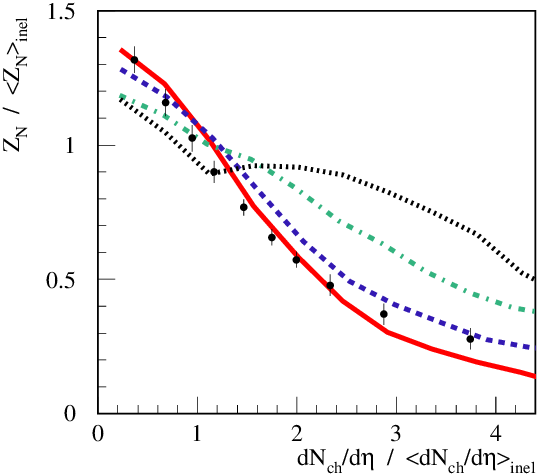}
\caption{Dependence of the relative strength of the ZDC signal, 
  $Z_N/ \langle  Z_N\rangle _{\rm inel>0}$,
   on the relative pseudorapidity
  density of charged hadrons   in the $|\eta | <0.5$ window, 
 for $pp$ collisions at $\sqrt{s}=13$ TeV,
    as calculated using different tunes of the QGSb model, 
   compared to ALICE data \cite{ach25} (points).
 The notations for the  lines are the same as in  Fig.\ \ref{fig:ppeta}.}
\label{fig:alice-corr}       
\end{figure}

However, a much more decisive discrimination
 can be made by studying correlations between  particle production  
  in the central rapidity range and in the proton fragmentation region 
  since this can allow one to test very basic model assumptions regarding 
  momentum distributions of constituent partons in the proton \cite{ost16}.
  Recently, such crucial measurements have been performed by the ALICE experiment at 
  the LHC, for $pp$ collisions at $\sqrt{s}=13$ TeV:
   investigating correlations beween  central pseudorapidity density
  of charged hadrons, $\left. dN_{\rm ch}/d\eta \right|_{\eta =0}$,
  and a signal strength $Z_N$ in   zero degree calorimeters (ZDC), the latter being dominated by the energy deposited in ZDC by
  secondary neutrons produced with $|\eta|>8.8$   \cite{ach22,ach25}. In Fig.\ \ref{fig:alice-corr},
these ALICE results are compared to the predictions of the above-discussed 
tunes 1, 2, 3, and 4 of QGSb,
regarding the dependence of the relative strength of the ZDC signal, 
  $Z_N/ \langle  Z_N\rangle _{\rm inel>0}$, on the relative pseudorapidity
  density of charged hadrons in the $|\eta | <0.5$ window, 
   $\left. \frac{dN_{\rm ch}}{d\eta} \right|_{|\eta| <0.5}
   / \langle \left. \frac{dN_{\rm ch}}{d\eta} \right|_{|\eta| <0.5}\rangle _{\rm inel>0}$.  
     Here $\langle  Z_N\rangle _{\rm inel>0}$ and 
   $\langle \left. \frac{dN_{\rm ch}}{d\eta} \right|_{|\eta| <0.5}\rangle _{\rm inel>0}$
   are the   average values of  $Z_N$ and
    $\left. \frac{dN_{\rm ch}}{d\eta} \right|_{|\eta| <0.5}$, respectively, 
    corresponding to the ``inel>0'' event selection of the ALICE experiment:
    at least one charged hadron in the   $|\eta | <1$ window.

    The strong anticorrelation between the   strength of the ZDC signal
    and the multiplicity of charged hadrons produced at  $|\eta | <0.5$,
    observed by ALICE, clearly supports  the Regge-inspired choice for the momentum
    distributions of string end partons,   
   $\alpha _{\rm sea} ^{\rm (sh)}=\alpha _{\rm sea} =\alpha _{\mathbb{R}}=0.5$,
 which is  adopted in the tunes 1 and 2, the corresponding results being shown by,
respectively,  solid and dashed lines in  Fig.\ \ref{fig:alice-corr}.
While the observed anticorrelation is well reproduced by both tunes, a somewhat
better overall agreement is provided by the default QGSb (tune 1)
characterized by a  larger value of  the overcriticality of the  ``semihard Pomeron'',  $\Delta_{\rm sh}=0.22$. For the tune 2, the smaller  $\Delta_{\rm sh}=0.21$ gives rise
to a  somewhat smaller multiple scattering rate at  $\sqrt{s}=13$ TeV, thereby 
underestimating the energy loss of leading neutrons. On the other hand, 
 for the high multiplicity
tail of the distribution of  $\left.  dN_{\rm ch}/d\eta \right|_{|\eta| <0.5}$,
the   strength of the ZDC signal predicted by the default QGSb is slightly smaller
than the measured one. Hence the data hint at the energy dependence of multiple
scattering rate (and of   $\sigma^{\rm tot}_{pp}$), which is intermediate between the
two tunes.

In contrast, considering softer  momentum
    distributions of string end partons, like in tunes 3 and 4, leads to a
    much weaker anticorrelation between central  hadron production
    and the forward neutron yield,
    thereby contradicting the experimental observations (cf.\ dash-dotted and 
    dotted lines  in  Fig.\ \ref{fig:alice-corr}). Consequently, the extreme modifications
    of the energy dependence of  $K^{\rm inel}_{p{\rm N}}$ and of $X_{\max}$,
    obtained using these tunes (dash-dotted and   dotted lines in Figs.\  \ref{fig:kinel}
    and  \ref{fig:xmax}), are disproved by the above-discussed ALICE measurements.
    These conclusions can be further cross checked and strengthened by studying 
    correlations between central and forward hadron production in proton-oxygen
    collisions, using experimental data collected during recent  proton-oxygen 
    run of the Large Hadron Collider.

\begin{figure}[htb]
\centering
\includegraphics[height=6.cm,width=0.48\textwidth]{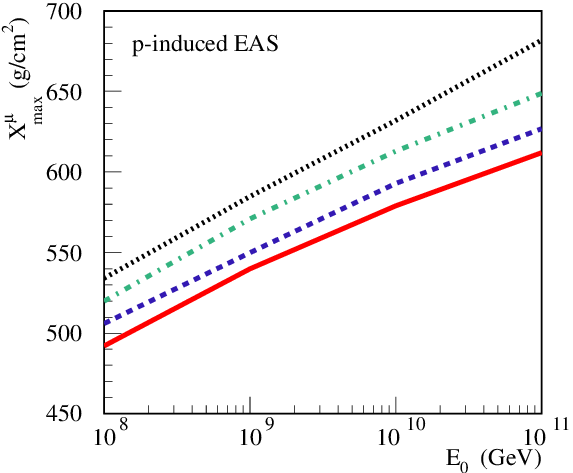}
\caption{Dependence of    maximal muon
production depth $X^{\mu}_{\max}$ ($E_{\mu}>1$ GeV),
for $p$-induced vertical EAS, on the   proton energy $E_0$.
 The notations for the  lines are the same as in  Fig.\ \ref{fig:ppeta}.}
\label{fig:xmumax}       
\end{figure}

 Additionally, as stressed in  \cite{ost24c,ost16a},
  a discrimination of predictions for the energy dependence of EAS
  maximum depth  can be performed by air shower experiments,
  based on  measurements of the  maximal muon
production depth $X^{\mu}_{\max}$ \cite{aab14}.
As one can see  in  Fig.\ \ref{fig:xmumax}, choosing softer distributions for 
constituent sea (anti)quarks attached to semihard Pomeron strings, one arrives
to substantially larger  $X^{\mu}_{\max}$ values predicted, compared to the default
QGSb: by up to $\simeq 40$
and $\simeq 70$ g/cm$^2$ for the tunes 3 and 4, respectively. Such a tendency
is not supported by the data of the Pierre Auger Observatory  \cite{aab14}:
to reach an agreement with the measured   maximal muon
production depth, 
one would have 
to postulate that UHECRs are much heavier than iron nuclei.

\subsubsection{How feasible is a smaller $X_{\max}$?  \label{xmax-low.sec}}
One can also study a possibility to shift the predicted shower maximum 
higher in the atmosphere, i.e., to have a smaller average $X_{\max}$.
In fact, there exists a theoretically well motivated mechanism leading
to a substantially higher inelasticity of proton-nucleus collisions at
high energies, related to the so-called ``diquark breaking'' in 
interacting nucleons \cite{kop89,cap96,kha96,dre05}. Up to now, we followed
the standard approach of the Quark-Gluon String and Dual Parton models
\cite{kai82,cap94}, assuming that the diquark of an interacting nucleon
survives multiple scattering as a coherent entity, i.e.,
remains in its color anti-triplet state. However, as rightfully argued
in \cite{kop89}, abundant multiple scattering should destroy the 
coherence between  the two valence quarks forming the diquark, transforming
the latter into color sextet state. As a consequence, the two quarks
should hadronize independently, being attached to two different
strings of color field, instead of having a single string coupled to the
diquark. The treatment in \cite{kop89,cap96,kha96} is based on the old idea
\cite{ros77} that the baryon content of a nucleon is not carried by its
valence quarks, being rather associated with the so-called string junction.
In the above-discussed scenario, the  string junction follows a slowed
down struck
valence quark, with the consequence that a leading nucleon is produced
with a substantially smaller fraction of initial light cone momentum
 ($E\pm p_z$) of the nucleon, compared to the case of diquark hadronization.
In \cite{dre05}, a somewhat different approach, though with similar
consequences, has been proposed. Namely, one considered a dense gluon
cloud incident on a nucleon, in the nucleon rest frame, such that 
 the two valence quarks forming the diquark suffer independent large
 transverse momentum ``kicks'' destroying the coherence of the diquark.
 
\begin{figure*}[t]
\centering
\includegraphics[height=5cm,width=0.9\textwidth]{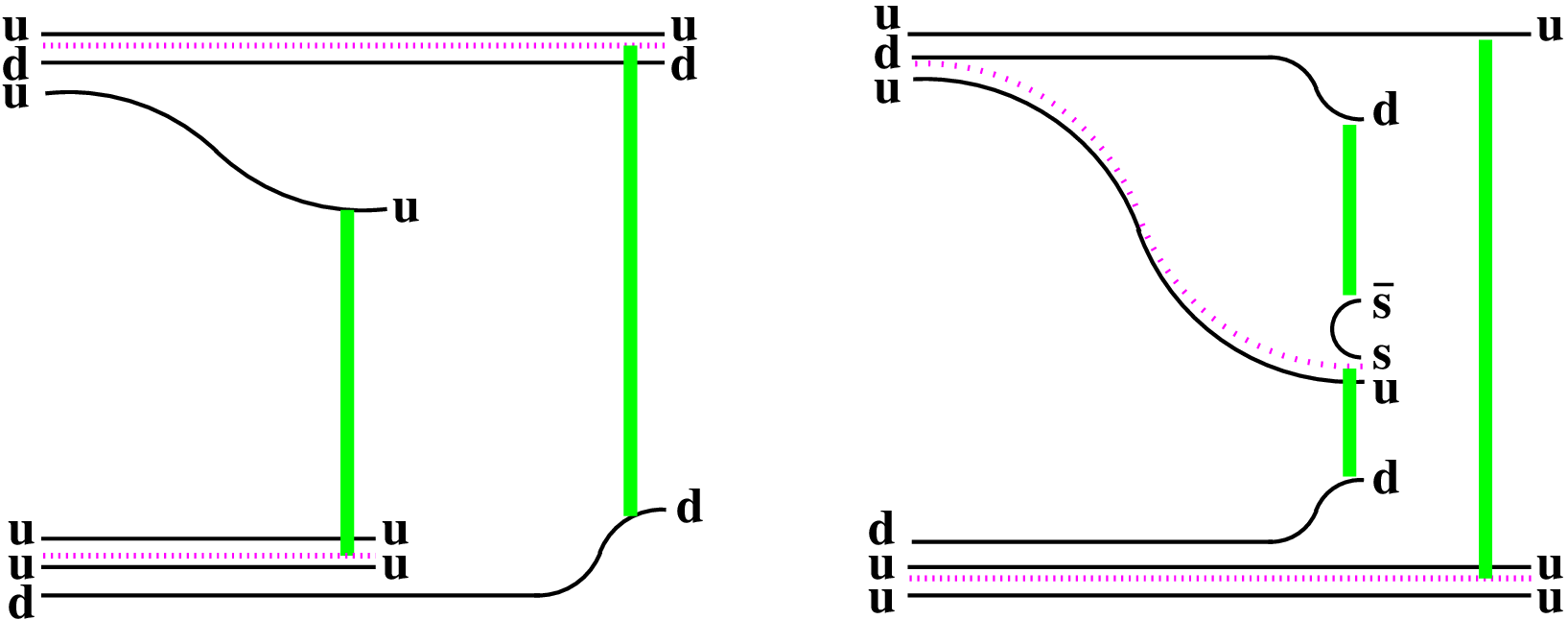}
\caption{Left: schematic view of the standard hadronization procedure for a
projectile proton. After the proton is split into a slowed down valence
quark (here $u$) and a diquark (here $ud$), the string junction shown by the
(upper) dotted magenta line remains coupled 
to the diquark and two strings of color field (thick vertical green lines)
 are stretched between these partons
and the corresponding constituents of the target (here $uu$ and $d$, respectively).
Right: a sketch of the diquark breaking mechanism for the
projectile proton. The string junction follows
the slowed down projectile valence quark ($u$), while the two remaining quarks ($d$ and $u$)
 hadronize independently. Upon creation of a quark-antiquark pair (here $\bar s s$) 
 from the vacuum, the created quark is coupled to the slowed valence quark and the junction
  to form a diquark (here $us$) and  three strings of color field are formed:
  here between $d$ and $\bar s$, $us$ and $d$, $u$ and $uu$.}
\label{fig:stj}       
\end{figure*}

 Here we choose to follow the approach of  \cite{kop89,cap96}, assuming that the
 first inelastic rescattering of a nucleon splits it into a valence quark
 and a diquark, whereas subsequent inelastic scattering processes may,
 with probability $w_{\rm stj}$, destroy the coherence of the diquark.
 Thus, for a nucleon undergoing $n$ inelastic rescatterings, i.e., being
 connected to $n$ cut Pomerons, with the probability $(1-w_{\rm stj})^{n-1}$
 the diquark survives as a coherent entity containing the string junction
 and the standard hadronization procedure exemplified in Fig.\ \ref{fig:stj}~(left)
 takes place. There,  two strings of color field are formed between 
  the (here projectile) quark and the diquark 
  [respectively, $u$ and $ud$ in  Fig.\ \ref{fig:stj}~(left)]
  and the corresponding constituent partons
    of the partner (here target) hadron (nucleus) [$uu$ and $d$ in  Fig.\ \ref{fig:stj}~(left)].
  In the other case depicted  in Fig.\ \ref{fig:stj}~(right), the string junction
  follows the struck valence quark, while the two remaining valence quarks
  hadronize independently, being each connected to its own string of color field.
  We then consider a creation of the quark-antiquark pair  from the vacuum 
  [$\bar s s$  in the example sketch of Fig.\ \ref{fig:stj}~(right)], 
  with the created quark being coupled to the slowed valence quark and the junction
  to form a diquark. In such a case, we have a formation of three strings:
  one stretched between one of the remaining valence quarks of the nucleon and
  the created antiquark [$d-\bar s$ in  Fig.\ \ref{fig:stj}~(right)], one between the created 
  diquark and a quark of the partner hadron [$us-d$ in Fig.\ \ref{fig:stj}~(right)],
  and the last string being formed between the remaining  valence quark
   of the nucleon and a constituent parton  of the partner hadron
    [$u-uu$ in  Fig.\ \ref{fig:stj}~(right)].
   
\begin{figure}[htb]
\centering
\includegraphics[height=6cm,width=0.48\textwidth]{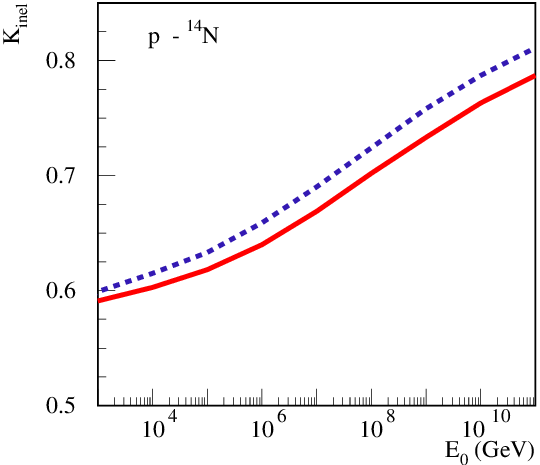}
\caption{Lab.\ energy dependence of the inelasticity $K^{\rm inel}_{p{\rm N}}$
of  proton-nitrogen collisions.
  The calculations performed using the default model (tune 1)
  or considering the diquark breaking mechanism  (tune 5) are shown, 
  respectively, by red solid  and blue dashed lines.}
 \label{fig:kinel-stj}       
\end{figure}

   We choose  $w_{\rm stj}=0.2$ and, to stay in agreement with measured proton
   spectra for $pp$ collisions at fixed target energies, use now $\alpha_N(0)=-0.25$
   for the intercept of the nucleon Regge trajectory\footnote{Interestingly, this value
   of  $\alpha_N(0)$ is in agreement with  results of standard Regge fits 
   (see, e.g., \cite{col77}).} (cf.\ Eq.\ (32) and Table 3 in \cite{ops26}). In addition,
   we switch off the special treatment of diquark hadronization in the case of single
   scattering (see Section 4.2 in  \cite{ops26}) -- by setting the corresponding
   parameter $\delta =0$. 
The results of the so obtained tune 5 of the model, regarding  the inelasticity of 
proton-nitrogen interactions, are plotted in   Fig.\ \ref{fig:kinel-stj},
where we now  have  a somewhat faster energy rise, compared to tune 1:
 with up to $\simeq 3$\% larger     $K^{\rm inel}_{p{\rm N}}$ at the highest energies.
 However, for the calculated average EAS maximum depth $X_{\max}$, we observed no
 noticeable differences from the default model predictions,\footnote{We verified by
 an explicit calculation that this remains the case when a higher diquark breaking
 probability, e.g.,  $w_{\rm stj}=0.5$, is used, provided other model parameters
  are properly adjusted to keep the agreement with measured proton
   spectra for $pp$ collisions at fixed target energies.} within the statistical
 uncertainties of the calculation ($\simeq 1-2$ g/cm$^2$), in contrast to drastic
 changes of  $X_{\max}$, predicted in \cite{dre05}. This seemingly paradoxical
 result can be understood if we realize that the considered mechanism has a sizable
 impact on the inelasticity of proton-nucleus collisions at not too large impact
 parameters $b$ only. There, copious multiple scattering already produces a sufficient
 ``stopping'' of leading nucleons and the additional ``softening'' of nucleon
 spectra is thus of weak importance for  the longitudinal EAS development.
 On the other hand, peripheral proton-air interactions at large $b$, which contribute
 to non-Gaussian tails of  $X_{\max}$ distributions and thereby influence significantly
 the average  $X_{\max}$, remain unaffected. In a sense, the obtained results may be
 regarded as an indication that the predictions of the default version of QGSb for
 the  average EAS maximum depth are already close to the corresponding lower bound.

\subsection{Aiming at a higher muon content  of  extensive air showers\label{nmu.sec}}
Let us now address the muon content of extensive air showers. In particular, we are
going to investigate possibilities to enlarge the predicted EAS muon number, $N_{\mu}$,
 in view of the reported excess of the number of muons in UHECR-induced air showers,
  compared to respective predictions of EAS simulations \cite{aab15,aab16} 
  (see also the review \cite{alb22} regarding the corresponding ``muon puzzle'').

As discussed in \cite{ost24d,hil97,rei21}, model predictions for  $N_{\mu}$ are 
closely correlated with the corresponding results for the moment
$\langle x_E^{\alpha_{\mu}} n_{\rm stable}^{\pi {\rm -air}}\rangle$ of the energy distribution
 of ``stable'' secondary hadrons  [(anti)nucleons, kaons, and charged pions] in
 pion-air interactions, defined as
 \begin{eqnarray}
 \langle x_E^{\alpha_{\mu}} n_{\rm stable}^{\pi {\rm -air}}(E_0)\rangle 
 \nonumber \\
 = \sum _{h={\rm stable}}  \int \!dx_E\; x_E^{\alpha_{\mu}}\,
 \frac{dn^{h}_{\pi {\rm -air}}(E_0,x_E)}{dx_E}\,.
 \label{eq:x*N}
\end{eqnarray}
The summation in the right-hand-side of Eq.\ (\ref{eq:x*N}) is performed
over  secondary hadrons having significant chances to interact in the atmosphere,
  instead of decaying\footnote{Obviously, such a definition of stable hadrons is energy dependent, e.g., $K^{\pm}$ can be regarded as being stable above $\sim 1$ TeV only.}, 
 $dn^{h}_{\pi {\rm -air}}/dx_E$ is their distribution, with respect to the 
energy fraction $x_E$ taken from the parent pion (upon decays of short-lived hadrons),
and $\alpha_{\mu}\simeq 0.9$   is the characteristic exponent for the
dependence of  $N_{\mu}$ on the primary energy. As is obvious from Eq.\ (\ref{eq:x*N}), 
$\langle x_E^{\alpha_{\mu}} n_{\rm stable}^{\pi {\rm -air}}\rangle$ is dominated by
forward production of stable hadrons in pion-air collisions, which is also the case
for   $N_{\mu}$ since more energetic secondaries initiate powerful enough subcascades
in the atmosphere and thereby contribute stronger to EAS muon content at ground.
On the other hand,   $N_{\mu}$ depends rather weakly on the energy dependence of the
multiplicity of pion-air interactions, in the very high energy limit, since the
energy rise of the multiplicity is driven  by hadron production in the
 central rapidity range, corresponding to small energy fractions  $x_E$,
 with the contributions of such hadrons to  Eq.\ (\ref{eq:x*N}) being
  damped\footnote{This is not the case for $\pi$-air interactions
at relatively low energies, $E_0 \lesssim 100$ GeV, where the central rapidity range
corresponds to relatively large $x_E \gtrsim 0.1$.} by the 
 factor $x_E^{\alpha_{\mu}}$.
 
Since   $\alpha_{\mu}$ is close to unity, the moment
 $\langle x_E^{\alpha_{\mu}} n_{\rm stable}^{\pi {\rm -air}}\rangle$ 
 can be approximated by the average
fraction of the parent pion energy   taken by all stable secondary
hadrons, i.e., by the case $\alpha_{\mu}=1$, which is approximately
equal to unity minus the energy fraction taken by neutral pions.
Therefore, to enhance the predicted  EAS muon content, one has to modify
the energy balance between stable secondary hadrons and neutral pions,
in favor of the former.

\begin{figure*}[p]
\centering
\includegraphics[height=5.4cm,width=0.9\textwidth]{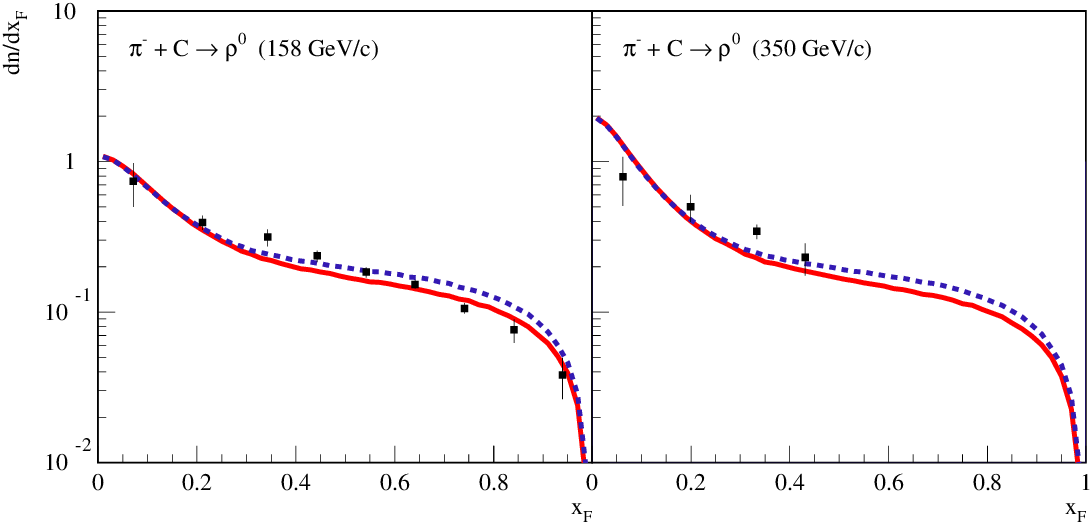}
\caption{Feynman $x$ distributions, $dn/dx_{\rm F}$, of $\rho^0$-mesons in c.m.\ frame,
 for $\pi^-C$ collisions at 158 GeV/c (left) and 350 GeV/c (right),
 as calculated using   tunes 1 and 6 of QGSb --
red solid and  blue dashed lines, respectively, compared to NA61/SHINE
  data \cite{adu17} (points).}
\label{fig:pimcrho}       
\end{figure*}

As discussed in \cite{ost13}, a particular mechanism
which efficiently reduces the energy ``leak'' into $\pi^0$ production
is the $t$-channel pion exchange  (see \cite{ops26} for the
details on the implementation of the process in the QGSb model).
This is because the process gives rise to forward production of $\rho$
mesons and, upon decays of the latter ($\rho^{\pm}\rightarrow \pi^{\pm}\pi^0$,
$\rho^{0}\rightarrow \pi^+\pi^-$), shifts the energy balance 
between secondary charged and neutral pions in favor of the former
(see the corresponding discussion in \cite{ost24d}).
In QGSb, the rate of  virtual pion emission by an incident pion
  is controlled by the
model parameter $g_{\pi /\pi}$, whose magnitude is seriously constrained 
by measurements of $\rho ^0$ meson production, notably,  by the NA61/SHINE 
experiment  \cite{adu17} (cf.\ solid lines in Fig.\ \ref{fig:pimcrho}).
\begin{figure*}[p]
\centering
\includegraphics[height=5.4cm,width=0.9\textwidth]{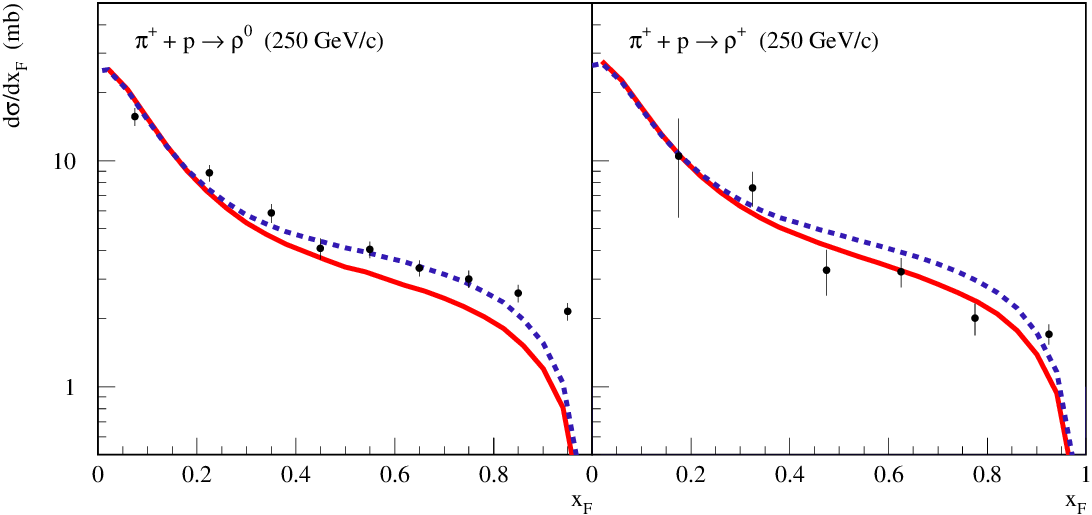}
\caption{$x_{\rm F}$-dependence of production cross sections in c.m.\ 
frame for $\rho^0$ (left) and $\rho^+$ (right) mesons, for $\pi^+p$ 
 collisions at 250 GeV/c, compared to EHS-NA22
 data \cite{aga90} (points).
 The notations for the  lines are the same as in  Fig.\ \ref{fig:pimcrho}.}
\label{fig:pirho250}       
\end{figure*}%
\begin{figure*}[p]
\centering
\includegraphics[height=5.4cm,width=0.9\textwidth]{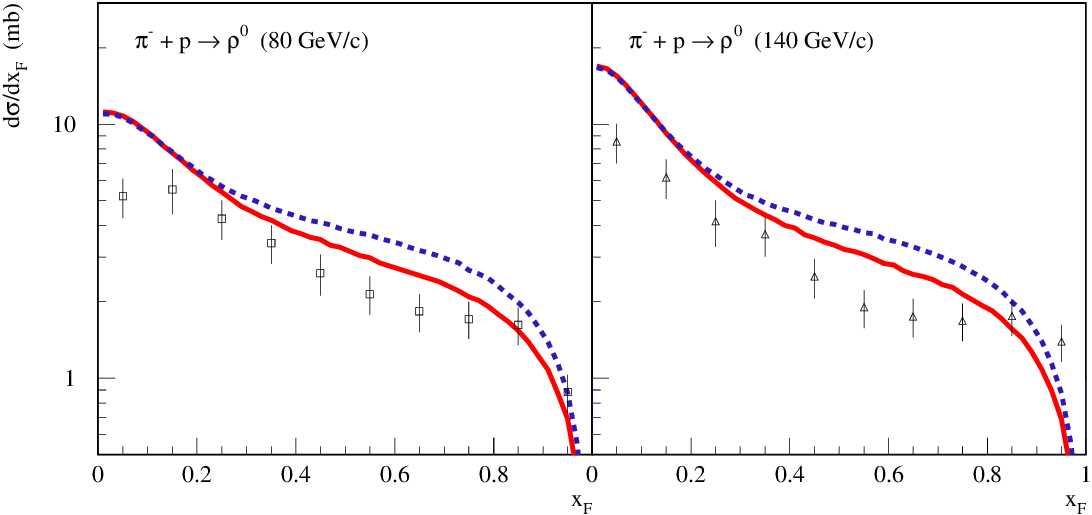}
\caption{$x_{\rm F}$-dependence of production cross sections in c.m.\ 
frame for $\rho^0$ mesons, for $\pi^-p$  collisions at 80 GeV/c (left)
and 140   GeV/c (right),  compared to the data of the OMEGA Photon
collaboration \cite{aps92} (points). 
 The notations for the  lines are the same as in  Fig.\ \ref{fig:pimcrho}.}
\label{fig:pimrho80}       
\end{figure*}%
 Here we can investigate the effect of a somewhat higher rate for the process,
 choosing a 50\% larger value for this parameter,  $g_{\pi /\pi}=0.225$.
  As one can see in
   Fig.\ \ref{fig:pirho250}, such a change (which we refer hereafter as tune 6)
   improves the agreement with the data of the EHS-NA22 experiment on forward 
    $\rho^0$ production, while leading to an overestimation of both  $\rho^+$
    yield reported by the same experiment and of   $\rho^0$ spectra, 
    compared to other measurements, notably, by  NA61/SHINE 
    (cf.\ Figs.\   \ref{fig:pimcrho},  \ref{fig:pirho250}, and \ref{fig:pimrho80}).
    Meanwhile, this gives rise to a very small change of the moment
    $\langle x_E^{\alpha_{\mu}} n_{\rm stable}^{\pi {\rm -air}}\rangle$ 
    (less than one per mille) and, correspondingly, to a minor enhancement
    ($\lesssim 1$\%) of  $N_{\mu}$, compared to the default QGSb results.

Alternatively, the EAS muon content can be enhanced by considering a more
abundant production of (anti)baryons and kaons in the projectile fragmentation region 
 in pion-air collisions (e.g., \cite{pie08,man23}), which contributes noticeably to
 $\langle x_E^{\alpha_{\mu}} n_{\rm stable}^{\pi {\rm -air}}\rangle$. 
 Those hadrons continue interacting in the atmosphere, thereby keeping the energy
more efficiently in the hadronic cascade, compared to pions, since decays of neutral pions
in the latter case work as an energy ``sink''.

As discussed in \cite{ops26}, forward production of kaons and (anti)nucleons in pion-nucleus interactions, in the QGSb model, was tuned to the corresponding results of the NA61/SHINE experiment \cite{adh23}. To improve the agreement with the data, we used a special treatment for the fragmentation of strings of color field, attached to  ``remnant''  (anti)quarks of the pion: choosing higher probabilities, compared to the 
standard string fragmentation procedure,
 for  strange  quark-antiquark pairs and  diquark-antidiquark pairs  creation from the vacuum, 
characterized by the parameters  $a_{s/\pi}$ and  $a_{ud/\pi}$, respectively (see Table 4 in \cite{ops26}).
  Phenomenologically, this can be interpreted
 as being related to   ``intrinsic strangeness'' and
    ``intrinsic baryonic content'' of pions.\footnote{A more theoretically rigorous approach
 would amount to consider $t$-channel exchanges  of, respectively,  $K^*$ and $\Delta$ Reggeons,  
  in the Reggeon-Reggeon-Pomeron ($\mathbb{RRP}$) configuration ($\mathbb{R}=K^*, \Delta$),
  similarly to the pion exchange case discussed above.} 
  
\begin{figure*}[p]
\centering
\includegraphics[height=8.cm,width=\textwidth]{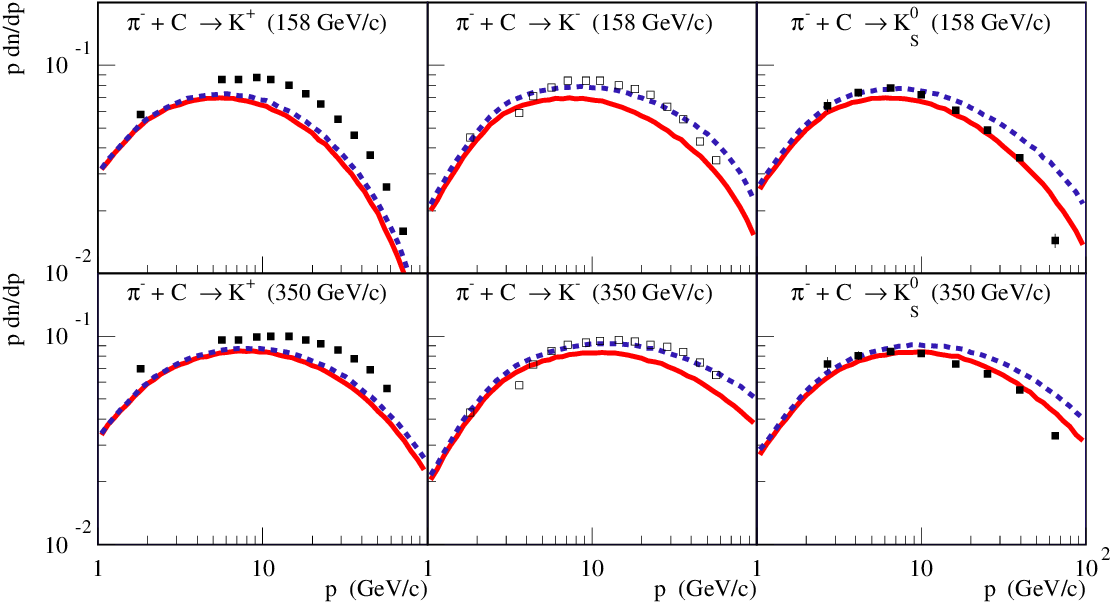}
\caption{Lab.\ momentum distributions of   $K^{+}$ (left panels),    $K^{-}$ (middle panels),
   and $K_{\rm S}^0$ (right panels) produced in $\pi^-$C collisions at 158 GeV/c (top row)
    and  350 GeV/c (bottom row), compared to NA61/SHINE data \cite{adh23} (points).
  The calculations performed using   tunes 1 and 7 of QGSb  are shown, 
  respectively, by red solid  and blue dashed lines.}
\label{fig:pimc158}       
\end{figure*}

\begin{figure}[htb]
\centering
\includegraphics[height=6cm,width=0.48\textwidth]{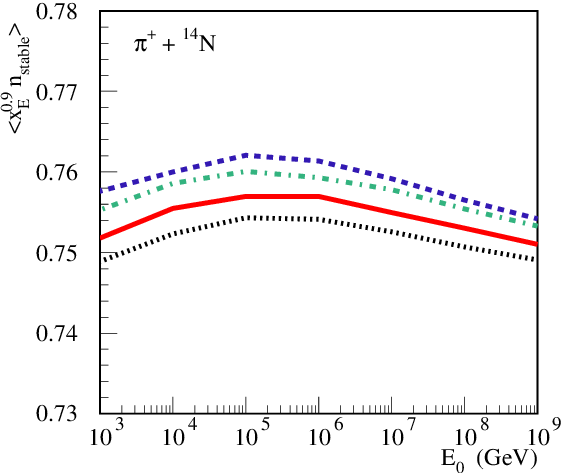}
\caption{Lab.\ energy dependence of the  moment
$\langle x_E^{0.9} n_{\rm stable}^{\pi {\rm -N}}\rangle$,
 as calculated using  the tunes 1, 7, 8, and 9 
of the QGSb model -- red solid,
   blue dashed, green  dash-dotted,  and black dotted lines, respectively.}
 \label{fig:nstable}       
\end{figure}

\begin{figure}[htb]
\centering
\includegraphics[height=6cm,width=0.48\textwidth]{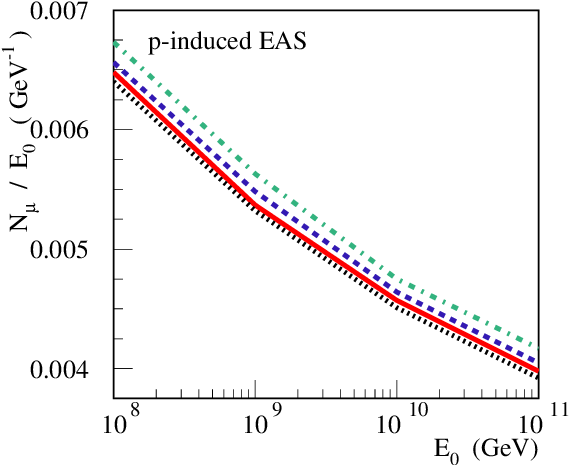}
\caption{Dependence on primary energy of   muon number  $N_{\mu}$ ($E_{\mu}>1$ GeV)
 at sea level, for proton-initiated vertical EAS.
 The notations for the  lines are the same as in  Fig.\ \ref{fig:nstable}.}
 \label{fig:nmu}       
\end{figure}

  While this allowed us to reach
  a reasonable agreement with the data of NA61/SHINE for  $K_{\rm S}^0$ production
  in  $\pi^-$C collisions at 158 and 350 GeV/c, the forward  
  yield of charged kaons remained underestimated, compared to the measurements
  (cf.\ solid lines in Fig.\ \ref{fig:pimc158}).
\begin{figure*}[p]
\centering
\includegraphics[height=9.5cm,width=\textwidth]{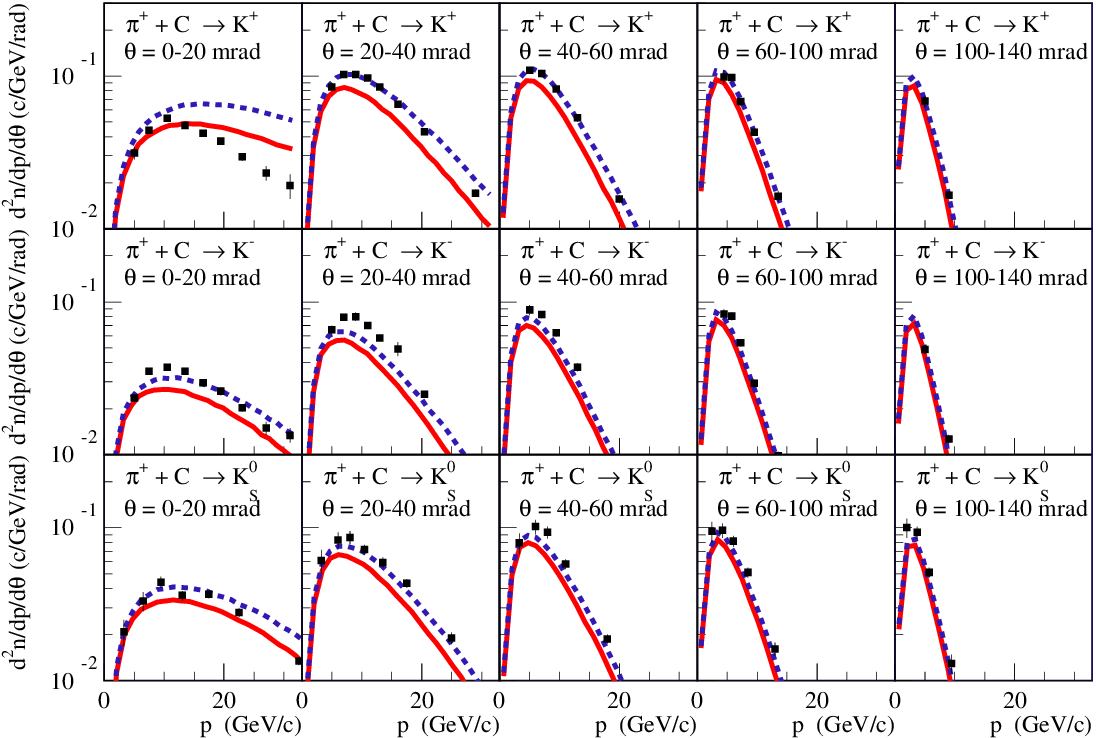}
\caption{Lab.\ momentum distributions of 
  $K^{+}$ (top row),    $K^{-}$ (middle row), and  $K_{\rm S}^0$ (bottom row)
   produced in $\pi^+$C collisions at 60 GeV/c, at different polar angles, as indicated in the plots,
 compared to NA61/SHINE data \cite{adu19a} (points).
 The notations for the  lines are the same as in  Fig.\ \ref{fig:pimc158}.}
\label{fig:pipc60}       
\end{figure*}%
  \begin{figure*}[t]
\centering
\includegraphics[height=6cm,width=0.9\textwidth]{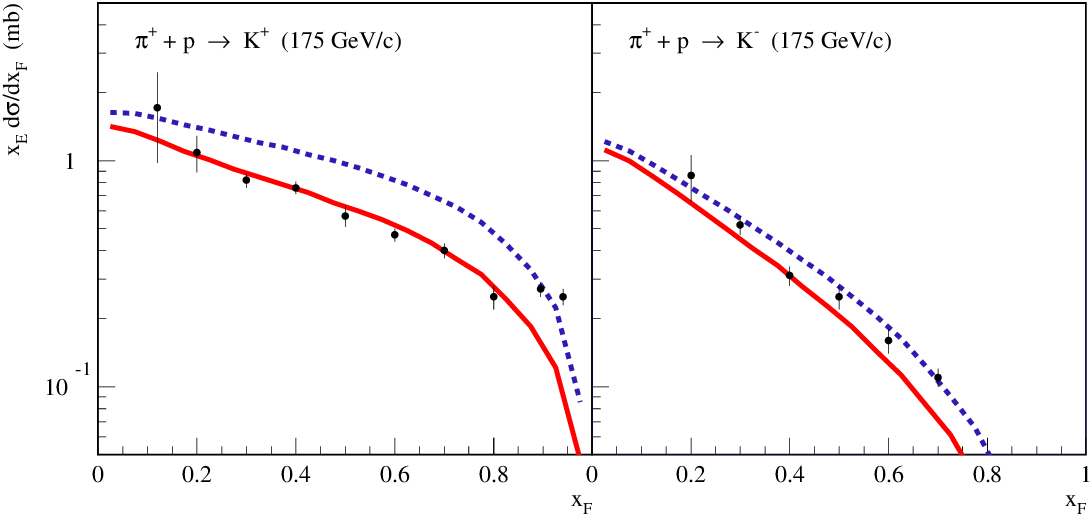}
\caption{$x_{\rm F}$-dependence of transverse momentum $p_t$ integrated
 invariant cross sections in c.m. frame, $x_E d\sigma /dx_{\rm F}$ ($x_E=2E/\sqrt{s}$), for
$K^+$ (left) and  $K^-$ (right) production in  $\pi^+p$ collisions at 175 GeV/c, 
compared to experimental data  \cite{bre82} (points).
 The notations for the  lines are the same as in  Fig.\ \ref{fig:pimc158}.}
\label{fig:pipp175}      
\end{figure*}%
 \begin{figure*}[htb]
\centering
\includegraphics[height=6cm,width=0.9\textwidth]{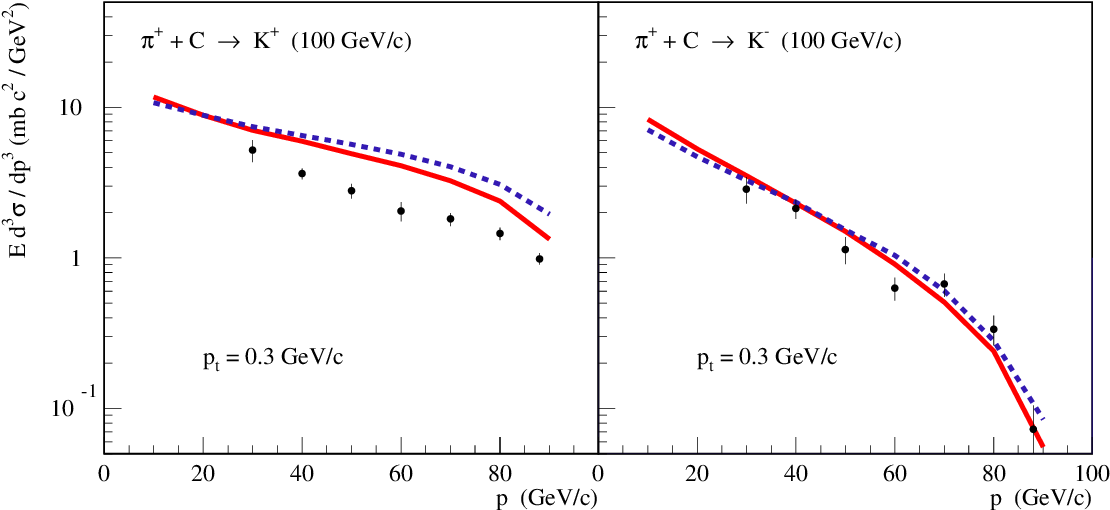}
\caption{Lab.\ momentum dependence of invariant cross sections, 
$E\, d^3\sigma/dp^3$, for   $p_t=0.3$ GeV/c,
for  $K^+$ (left) and  $K^-$ (right)  production in
 $\pi^+$C collisions at 100 GeV/c, 
compared to experimental data  \cite{bar83} (points).
 The notations for the  lines are the same as in  Fig.\ \ref{fig:pimc158}.}
\label{fig:pipc100}      
\end{figure*}%
  Leaving aside the question regarding a hypothetical 
   isospin symmetry violation  \cite{adh25}, we can study here the impact on $N_{\mu}$ of a more
   abundant forward kaon production, using a factor two larger value 
   $a_{s/\pi}=0.4$, while keeping the default values for all other parameters of the model.
 As one can see  in Fig.\ \ref{fig:pimc158}, for the so-obtained tune 7 of the model,
  we reach a reasonably good agreement with the measured $K^-$ spectra, while still
  largely underestimating  $K^+$ production and overestimating now the forward yield of   $K_{\rm S}^0$. Regarding kaon production in pion-carbon collisions at  a smaller incident pion momentum, 60  GeV/c, the  agreement with the data is improved both for charged and for neutral kaons, as one can see in Fig.\ \ref{fig:pipc60}. However,  for very forward production of  $K^+$ at small polar angles, $\theta <20$ mrad,
 the calculated  spectra are substantially harder than
 the measured ones. On the other hand, the obtained model tune
 largely overestimates forward kaon production, compared to results of other experiments,
 both for kaon-proton and kaon-nucleus collisions, see Figs.\  \ref{fig:pipp175} and  \ref{fig:pipc100}.
 For the moment 
 $\langle x_E^{\alpha_{\mu}} n_{\rm stable}^{\pi {\rm -N}}\rangle$, $\alpha_{\mu}=0.9$,
 for pion-nitrogen collisions, we obtain with this tune an enhancement,
  relative to the default QGSb, which ranges between  0.8\% at $E_0=1$ TeV and  
  0.4\%  at the highest energies (cf.\ solid and dashed lines
  in  Fig.\ \ref{fig:nstable}). 
  For the muon number  $N_{\mu}$ ($E_{\mu}>1$ GeV) at sea level, 
  for vertical proton-induced air showers, this leads to an enhancement of up to
 $\simeq 2$\%,  compared to the default QGSb results, as one can see  in Fig.\  \ref{fig:nmu}.

\begin{figure*}[t]
\centering
\includegraphics[height=11.cm,width=\textwidth]{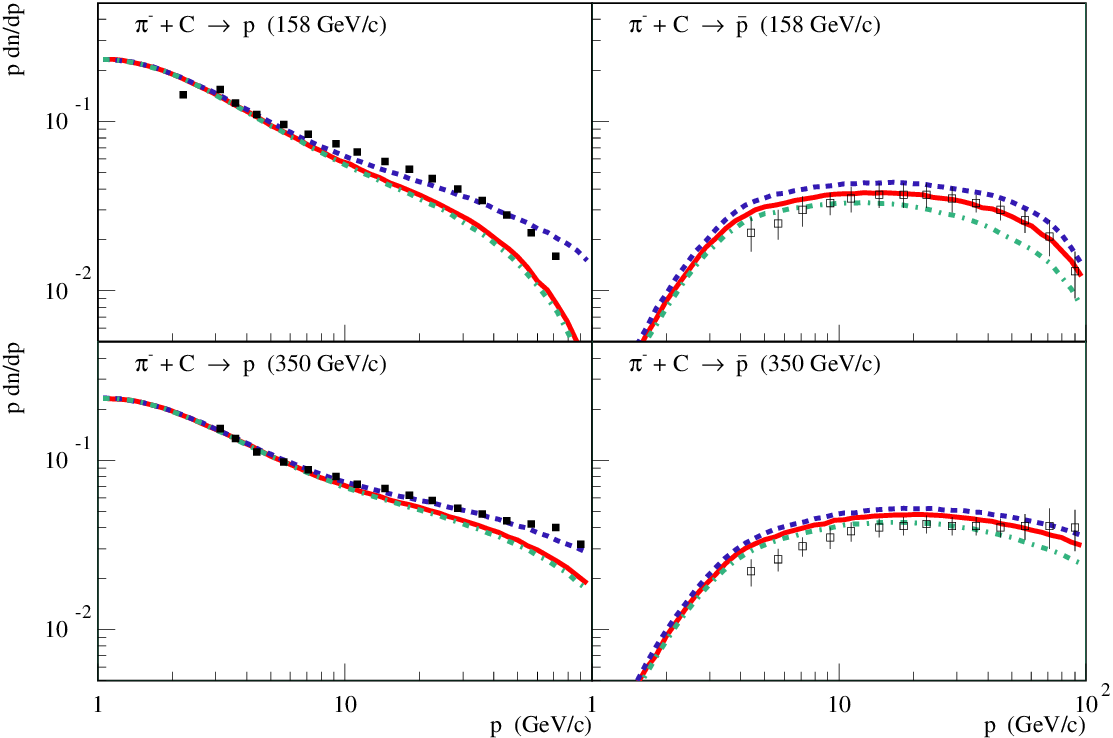}
\caption{Lab.\ momentum distributions of   protons (left panels)  and antiprotons (right panels), produced in $\pi^-$C collisions at 158 GeV/c (top row)
    and  350 GeV/c (bottom row), compared to NA61/SHINE data \cite{adh23} (points).
  The calculations performed using the tunes 1, 8, and 9 of QGSb  are shown, 
  respectively, by red solid, blue dashed, and green dash-dotted  lines.}
\label{fig:pimc158-p}       
\end{figure*}

\begin{figure*}[t]
\centering
\includegraphics[height=3.cm,width=\textwidth]{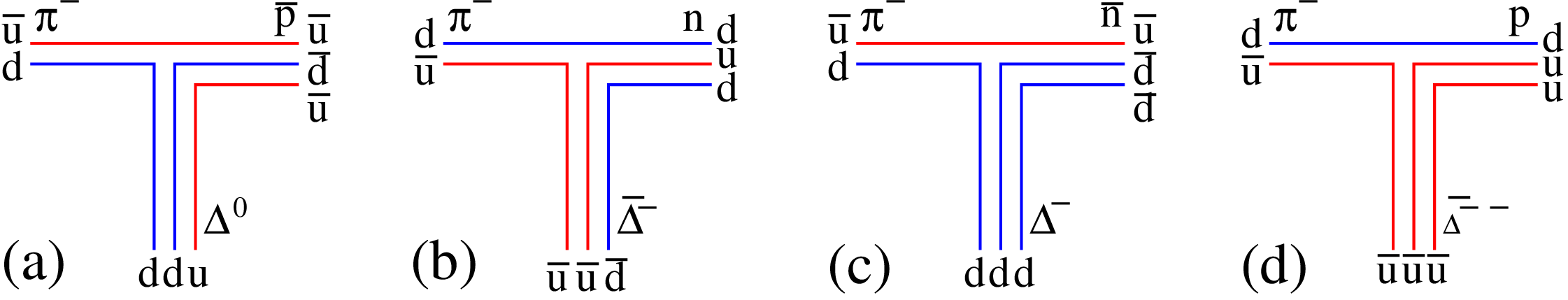}
\caption{Schematic view of the contributions of $t$-channel exchange of $\Delta$-Reggeons to forward
(anti)nucleon production in pion-proton and pion-nucleus collisions. (a) The $ud$ diquark
from a vacuum-created  $ud-\bar u \bar d$ pair couples with the slowed down valence quark $d$,
forming a virtual  $\Delta^0$ baryon which subsequently interacts with the target (not shown in the
 Figure). The  $\bar u \bar d$ antidiquark  is combined with the ``remnant'' valence antiquark $\bar u$
 to form a leading antiproton: $\pi^-\stackrel{(\Delta^0)}{\rightarrow}\bar p$. (b)~Same as in (a)
 but for an emission of a virtual  $\bar \Delta^-$ baryon and a formation of a leading neutron:
 $\pi^-\stackrel{(\bar \Delta^-)}{\rightarrow}n$. The same mechanism proceeds in (c) and (d)
 via a creation of, respectively,  $dd-\bar d \bar d$  and $uu-\bar u \bar u$ pairs from the vacuum:
  $\pi^-\stackrel{(\Delta^-)}{\rightarrow}\bar n$   and 
 $\pi^-\stackrel{(\bar \Delta^{--})}{\rightarrow}p$.  }
\label{fig:delta-ex}       
\end{figure*}

   Let us now investigate the impact on EAS muon content of an enhancement of (anti)nucleon
   production in pion-air interactions. For the default QGSb, we have a reasonable agreement with 
    NA61/SHINE measurements of antiproton spectra in $\pi^-$C collisions at 158 and 350 GeV/c,
    while the calculated forward production of protons is substantially smaller than observed
    by the NA61/SHINE experiment   (see solid lines in Fig.\  \ref{fig:pimc158-p}).
 As already mentioned   above, we employed a special treatment for the hadronization of ``remnant'' (anti)quarks of the
    pion, using   a higher probability for creation of $ud-\bar u \bar d$ pairs from the
     vacuum, governed by the parameter  $a_{ud/\pi}$, compared to the corresponding parameter $a_{ud}$
     of the standard string fragmentation procedure (cf.\ Table 4 in \cite{ops26}).
     Qualitatively, such a prescription can be justified by an analogy to a more rigorous Regge
     treatment involving an emission of a virtual  $\Delta$ (anti)baryon by the incident pion, the
     corresponding contributions being shown schematically in Fig.\ \ref{fig:delta-ex}~(a, b),
     for the particular case of an incident  $\pi^-$. In  Fig.\ \ref{fig:delta-ex}~(a), upon emission of
    a   $\Delta ^0$  baryon, the  projectile $\pi^-$ converts into an antiproton, 
$\pi^-\stackrel{(\Delta^0)}{\rightarrow}\bar p$,  while the virtual $\Delta^0$ interacts with the target
(not shown in the Figure). Likewise, in  Fig.\ \ref{fig:delta-ex}~(b), one has an emission of a virtual
$\bar \Delta ^-$, thereby converting the incident pion into a neutron:
 $\pi^-\stackrel{(\bar \Delta^-)}{\rightarrow}n$. Here, pursuing further the analogy to the Regge
 treatment and invoking the isospin symmetry considerations, one would have to include, with equal
 probabilities, the processes shown in  Fig.\ \ref{fig:delta-ex}~(c, d), corresponding
  to emissions of  $\Delta^-$ and $\bar \Delta^{--}$  by the incident pion. 
  In our phenomenological approach, this would amount considering a creation of 
   $dd-\bar d \bar d$  and $uu-\bar u \bar u$ pairs from the   vacuum, with the same probabilities
   as for   $ud-\bar u \bar d$ pairs, in the hadronization  of ``remnant'' (anti)quarks of the
    pion.
    
\begin{figure*}[t]
\centering
\includegraphics[height=9cm,width=\textwidth]{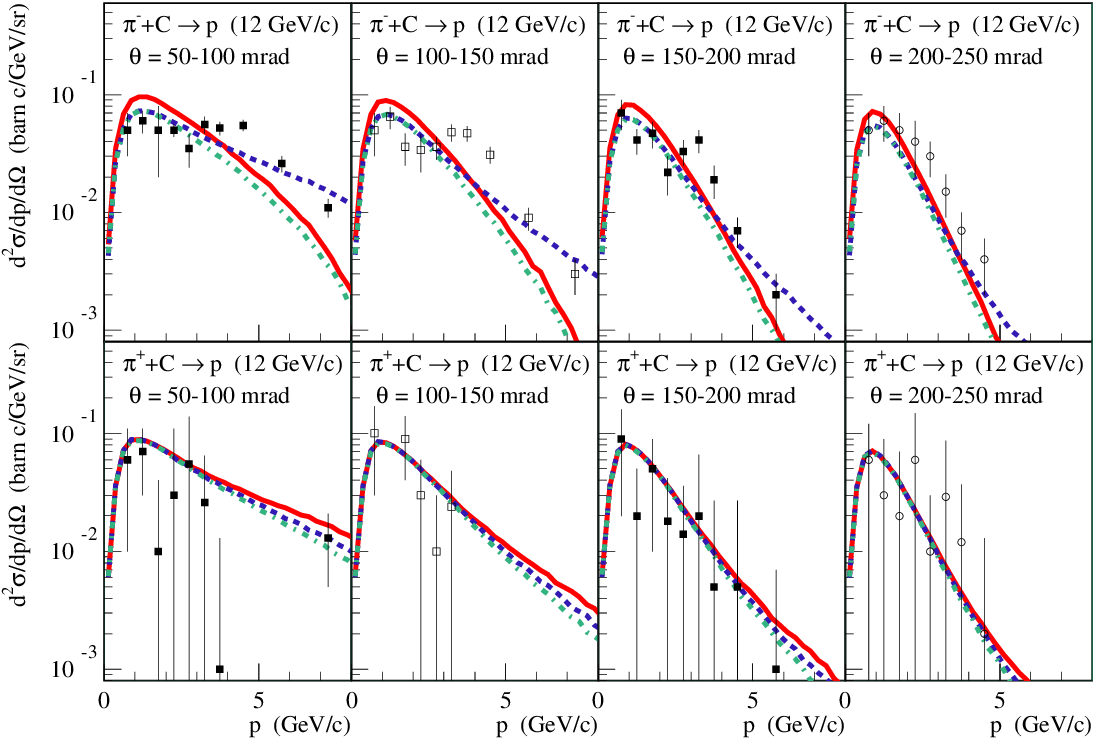}
\caption{Lab.\ momentum  dependence of inclusive cross sections for 
 proton production, at  different polar angles,  as indicated in the plots,  in  $\pi^-$C (top) and
 $\pi^+$C (bottom) collisions at  12 GeV/c,
compared to HARP data \cite{apo10} (points).
 The notations for the  lines are the same as in  Fig.\ \ref{fig:pimc158-p}.}
\label{fig:pipcharp}       
\end{figure*}

\begin{figure*}[t]
\centering
\includegraphics[height=6cm,width=0.9\textwidth]{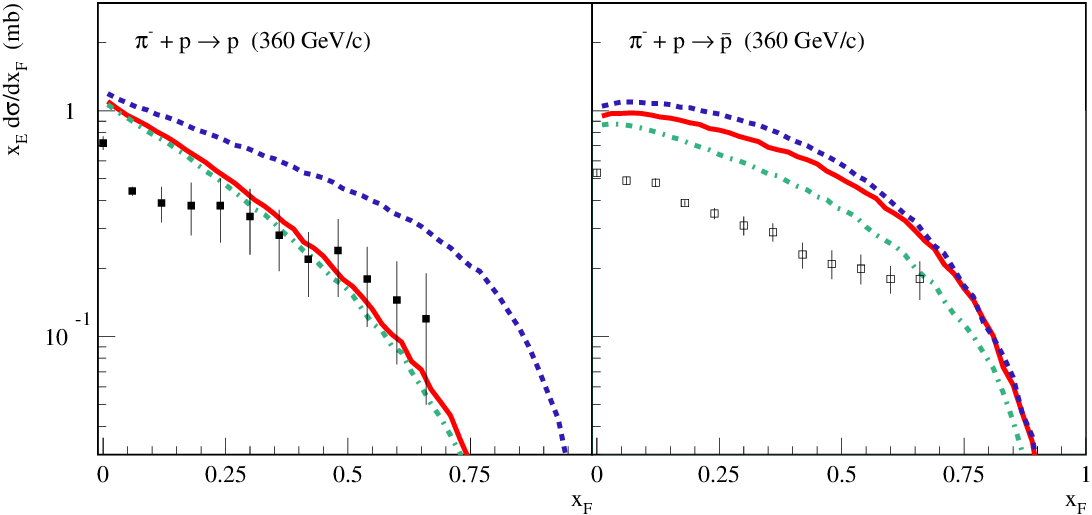}
\caption{$x_{\rm F}$-dependence of $p_t$-integrated
 invariant cross sections in c.m. frame, $x_E d\sigma /dx_{\rm F}$, for
  proton (left) and antiproton (right) production in $\pi^-p$ 
 collisions at 360 GeV/c, compared to LEBC-EHS data \cite{agu87} (points). 
 The notations for the  lines are the same as in  Fig.\ \ref{fig:pimc158-p}.}
\label{fig:pimp360-p}       
\end{figure*}

    We study such a modification, using   $a_{dd/\pi}=a_{uu/\pi}=a_{ud/\pi}=0.15$, which we refer 
    hereafter as tune 8. For the spectra of protons and antiprotons in   $\pi^-$C collisions 
    at 158 and 350 GeV/c, plotted as dashed lines in  Fig.\  \ref{fig:pimc158-p},
     we now have a reasonable agreement with  NA61/SHINE data. Regarding proton production in pion-carbon collisions at a smaller momentum of the incident pion,
 12 GeV/c, the discussed modification  improves slightly the agreement with  
measurements  of the HARP experiment \cite{apo10}, 
 as one can see in Fig.\   \ref{fig:pipcharp}.
  However,      for $p$ and $\bar p$ production in   $\pi^-p$ interactions at 360 GeV/c, we arrive with this
      tune to a factor two excess, compared to the corresponding data
    of the LEBC-EHS experiment \cite{agu87},   see     Fig.\  \ref{fig:pimp360-p}.
 For the moment $\langle x_E^{\alpha_{\mu}} n_{\rm stable}^{\pi {\rm -N}}\rangle$, 
$\alpha_{\mu}=0.9$, the enhancement obtained with this model tune is comparable 
to the one we had for tune~7 (with enhanced kaon production):
ranging between 0.5\%  and 0.3\%, the corresponding energy dependence being
plotted in  Fig.\  \ref{fig:nstable} as   dash-dotted line.
The calculated EAS muon number $N_{\mu}$  exceeds the corresponding
results of the default model by up to 5\%,  as one can see  in Fig.\  \ref{fig:nmu}.

Finally, we can try the opposite changes: switching down the special treatment 
of the hadronization of ``remnant'' (anti)quarks of the pion, i.e., restricting
ourselves with the creation of  $ud-\bar u \bar d$ diquark-antidiquark pairs
from the vacuum and using one and the same value for the corresponding model parameters,
for fragmentation of all strings (setting   $a_{ud/\pi}=a_{ud}=0.1$).
In such a case (referred to as tune 9), we marginally improve the agreement
 with  LEBC-EHS data on $p$ and $\bar p$ production 
 (cf.\ dash-dotted lines in Fig.\  \ref{fig:pimp360-p}), while
 further moving away from the corresponding measurements by the 
 NA61/SHINE and HARP experiments, as one can see in
  Figs.\  \ref{fig:pimc158-p}  and  \ref{fig:pipcharp}.
 For this tune, the calculated moment
 $\langle x_E^{0.9} n_{\rm stable}^{\pi {\rm -N}}\rangle$
 is smaller than for the default QGSb by $\simeq 0.3$\% (cf.\ dotted line in
 Fig.\  \ref{fig:nstable}) while the obtained number
 of muons is decreased by   $1-2$\%, 
 see  Fig.\  \ref{fig:nmu}.
 
 For all the alternative model tunes studied in this Section, the multiplicity of secondary hadrons produced in pion-air collisions remained practically unchanged, the differences to the default model results being at subpercent level. 
 Hence, the obtained modifications of EAS muon content were caused by changes of forward spectra of secondary hadrons. 
 Here, the largest impact on $N_{\mu}$ was observed when modifying  forward production of (anti)nucleons in pion-air interactions. 
 Partly this is because the corresponding treatment is not sufficiently constrained by relevant accelerator data, due to serious contradictions between different measurements of (anti)proton yields in the fragmentation region of incident pion, in $\pi p$ and $\pi A$ collisions.
 
 For convenience, we compile in Table \ref{Flo:param} the parameter changes, for the considered alternative tunes of the model.
  \begin{table*}[t]
\begin{centering}
\begin{tabular}{|llllllllllll|}
\hline 
   &    $\Delta _{\rm sh}$ &  $\alpha_{\rm sea}^{\rm (sh)}$ & 
    $\Lambda^{\rm (sh)}$ &   $\alpha_N(0)$ &   $\delta$ &   $g_{\pi /\pi}$ &
  $a_{s /\pi}$ &   $a_{ud}$  &  $a_{ud/\pi}$ &  $a_{uu/\pi}=a_{dd/\pi}$ & $w_{\rm stj}$
  \tabularnewline
\hline 
default value  & 0.22 & 0.5 & 3.6 & -0.05 & 1 & 0.15  & 0.1  & 0.1 & 0.19 &  0 & 0
\tabularnewline
tune 2   & 0.21 &   & 4.6 &  &  &   &    &   &   &   &
\tabularnewline
tune 3   & 0.21 & 0.8   & 6.5 &  &  &   &    &   &   &   &
\tabularnewline
tune 4   & 0.21 & 0.95  & 9.5 &  &  &   &    &   &   &   &
\tabularnewline
tune 5   &      &       &     & -0.25  & 0  &   &    &   &   &   & 0.2
\tabularnewline
tune 6   &      &       &     &     &    & 0.225  &    &   &   &   & 
\tabularnewline
tune 7   &      &       &     &     &    &   & 0.4   &   &   &   & 
\tabularnewline
tune 8   &      &       &     &     &    &   &    &   &  0.15 & 0.15   & 
\tabularnewline
tune 9   &      &       &     &     &    &   &    &   &  0.1 &     & 
\tabularnewline
\hline
\end{tabular}\caption{Summary of the parameter changes, for the  alternative tunes of the QGSb model.}
\label{Flo:param}
\par\end{centering}
 \end{table*}

\section{Conclusions \label{concl.sec}}
We employed the new Monte Carlo generator of hadronic collisions, QGSb, 
for studying model uncertainties for calculated characteristics of
 extensive air showers   initiated by interactions of high energy cosmic rays in the atmosphere. We specifically concentrated on modifying predictions for two  key EAS observables mostly used in experimental studies of nuclear 
 composition of high energy cosmic rays: the shower maximum depth $X_{\max}$ and the muon number at ground level $N_{\mu}$. In particular, we discussed in some detail the relevant 
  physics mechanisms and the corresponding model parameters, studying also the
  impact of the considered modifications of the model on the level of 
  agreement of its predictions with measured  characteristics of hadron-proton
  and hadron-nucleus collisions.
  
  Regarding EAS maximum depth, our main interest was to investigate possibilities to predict a slower air shower development, which could potentially improve the agreement with experimental data of the Pierre Auger Observatory. We demonstrated that the most efficient way for obtaining larger  $X_{\max}$ values is to consider softer momentum distributions for hadron constituents to which strings of color field are attached, in agreement with previous studies. However, the most extreme modifications
  of the kind, studied by us, proved to be incompatible with recent results of the ALICE
  experiment at LHC, regarding correlations between secondary hadron production at central
  rapidities and in the proton fragmentation region, in proton-proton collisions. Moreover,
 those modifications are strongly disfavored by the data of the  Pierre Auger Observatory,
   regarding the maximal muon production depth of UHECR-induced EAS: since a compatibility
    with the data can only be achieved for primary nuclei heavier than iron.
    On the other hand, we have not found much freedom for speeding up the hadronic cascade 
    development, i.e., for obtaining sizably smaller $X_{\max}$ values. In particular, contrary
    to what has been claimed in the literature, implementing the so-called diquark 
    splitting mechanism does not produce any appreciable change of the predicted $X_{\max}$,
    provided other relevant model parameters are adjusted to keep an agreement with measured
    proton spectra in $pp$ collisions at fixed target energies.
  
  Further, in relation to the so-called muon puzzle, we investigated  possibilities to enlarge the predicted EAS muon content. The latter was shown to depend strongly on the treatment of forward hadron production, notably, on the energy balance between (relatively) stable secondary hadrons  and neutral pions. 
The largest  enhancement of  $N_{\mu}$, up to $\simeq 5$\%, was obtained 
 by  considering harder spectra of  (anti)nucleons produced in pion-air collisions.
  A reduction of the   corresponding uncertainty
   is currently problematic because of serious contradictions between different measurements of 
  (anti)proton production in pion-proton and pion-nucleus interactions.
  
  While the studied modifications of the model are not comprehensive,
  we clearly outlined the main directions for an alternative  tuning
 of model parameters and identified relevant accelerator data constraining
 such changes. On the other hand,
further modifications of   EAS   predictions can potentially be caused by other interaction mechanisms not considered in this work, e.g., by collective effects in the hadronization procedure \cite{bau23}.
  Such mechanisms may be implemented in the model at a later stage.

\subsection*{Acknowledgments}
S.O.\ acknowledges   support from  Deutsche Forschungsgemeinschaft 
(project  550225003). 
 G.S.\ acknowledges
support by the Bundesministerium f\"ur Bildung
und Forschung, under grant  05A23GU3, and
by the Deutsche Forschungsgemeinschaft  under    
  Germany's Excellence Strategy -- EXC 2121 ``Quantum Universe'' -- 390833306.


\begin{thebibliography}{99}

  \bibitem{nag00}
  M.\ Nagano and A.\ A.\ Watson, 
  {\em Observations and implications of the ultrahigh-energy cosmic rays}, 
   Rev.\ Mod.\ Phys.\ {\bf 72}, 689 (2000).
   
  \bibitem{blu09}  
  J.\ Blumer, R.\ Engel, and  J.\ R.\ Horandel,
{\em Cosmic rays from the knee to the highest energies}, 
        Prog.\ Part.\ Nucl.\ Phys.\  {\bf  63}, 293 (2009).

\bibitem{hec98}
D.\ Heck, J.\ Knapp, J.\ N.\ Capdevielle, G.\ Schatz, and T.\ Thouw, 
{\em CORSIKA: A Monte Carlo
code to simulate extensive air showers},
 Forschungszentrum Karlsruhe Internal Report FZKA-6019 (1998).

\bibitem{eng11}
 R.\ Engel, D.\ Heck, and T.\ Pierog, 
{\em  Extensive air showers and hadronic interactions at high energy}, 
 Ann.\ Rev.\ Nucl.\ Part.\ Sci.\  {\bf 61}, 467 (2011).

\bibitem{ulr11}
R.\  Ulrich, R.\ Engel, and M.\ Unger,
{\em  Hadronic multiparticle production at ultra-high energies and extensive air showers},
Phys.\ Rev.\  D  {\bf 83}, 054026 (2011).  

  \bibitem{par11}
    R.\ D.\ Parsons, C.\  Bleve, S.\ S.\  Ostapchenko, and J.\  Knapp,
    {\em   Systematic uncertainties in air shower measurements from 
    high-energy hadronic interaction models},
    Astropart.\ Phys.\   {\bf 34},  832 (2011).

   \bibitem{ost24c} S.\ Ostapchenko and G.\ Sigl, 
   {\em Model uncertainties for the predicted maximum depth of extensive air showers},
 Phys.\  Rev.\ D {\bf  110},   063041   (2024).  

   \bibitem{ost24d} S.\ Ostapchenko and G.\ Sigl, 
{\em On the model uncertainties for the predicted muon content
 of extensive air showers},
   Astropart.\ Phys.\    {\bf 163}, 103004 (2024).

 \bibitem{ost16}
 S.\ Ostapchenko, M.\ Bleicher, T.\ Pierog, and K.\ Werner, 
   {\em Constraining high energy interaction
   mechanisms by studying forward hadron production at the LHC}, 
  Phys.\ Rev.\  D    {\bf 94},   114026 (2016).
 
 \bibitem{ost23}
S.~Ostapchenko, 
{\em Cosmic ray interactions in the atmosphere: QGSJET-III and other models},
SciPost Phys.\ Proc.\  {\bf 13}, 004 (2023).

  \bibitem{ops25}
  S.\ Ostapchenko, T.\ Pierog, and G.\ Sigl,
{\em   A new approach to modeling cosmic ray interactions},
PoS (ICRC2025),  351  (2025).

  \bibitem{ops26}
  S.\ Ostapchenko, T.\ Pierog, and G.\ Sigl,
   {\em Basic model for high energy cosmic ray interactions}, 
  Phys.\ Rev.\  D    {\bf 113},  074001 (2026).

 \bibitem{abd24}
 A.\  Abdul Halim  et al.\ (Pierre Auger Collaboration), 
{\em Testing hadronic-model predictions of depth of maximum of air-shower profiles and 
ground-particle signals using hybrid data of the Pierre Auger Observatory},
 Phys.\  Rev.\ D {\bf 109},   102001    (2024).  

  \bibitem{gla56}
 R.\ J.\ Glauber, 
 {\em  High-energy collision theory},
 In: Lectures in theoretical physics, ed.\ by W.~E.~Brittin 
and L.\ G.\ Dunham, Interscience Publishers (New York, 1959), vol.\ 1,
 pp.\ 315-414.
  
\bibitem{gri69}
 V.\ N.\ Gribov, 
 {\em Glauber corrections and the interaction between high-energy 
hadrons and nuclei},
  Sov.\ Phys.\ JETP  {\bf 29}, 483, (1969). 

\bibitem{ant19}
 G.\ Antchev et al. (TOTEM Collaboration),
{\em First measurement of elastic, inelastic and total cross-section 
at  $\sqrt{s}=13$ TeV by TOTEM and overview of cross-section data at LHC energies},	
 Eur.\ Phys.\ J.\  C \textbf{79},  103 (2019).

\bibitem{aad23}
G.\ Aad  et al.\ (ATLAS Collaboration),
  {\em  Measurement of the total cross section and $\rho$-parameter from elastic scattering in pp collisions at $\sqrt{s}=13$ TeV with the ATLAS detector},
 Eur.~Phys.~J.~C {\bf 83},  441  (2023).

	\bibitem{pdg}
 R.\ L.\ Workman  {\em et al.} (Particle Data Group), 
 {\em Review of Particle Physics},
   Prog.\ Theor.\ Exp.\ Phys.\  {\bf 2022}, 083C01 (2022).

\bibitem{cha14}
S.\ Chatrchyan et al.\ (CMS and TOTEM Collaborations),
{\em  Measurement of pseudorapidity distributions of charged particles in
proton-proton collisions at  $\sqrt{s}=8$ TeV by
the CMS and TOTEM experiments},	
 Eur.~Phys.~J.~C {\bf 74},  3053  (2014).

 \bibitem{ost03}
 S.\ S.\ Ostapchenko,
 {\em Contemporary models of high-energy interactions:
  Present status and perspectives},
  J.\ Phys.\ G {\bf 29}, 831 (2003).

   \bibitem{adr15} 
  O.\ Adriani   {\em et al.} (LHCf Collaboration),
  {\em Measurement of very forward neutron energy spectra for    $\sqrt{s}=7$ TeV
   proton-proton collisions at the Large Hadron Collider},
  Phys.~Lett.~B   {\bf 750},  360 (2015).

   \bibitem{adr18} 
  O.\ Adriani   {\em et al.} (LHCf Collaboration),
  {\em Measurement of inclusive forward neutron production cross 
   section in proton-proton collisions at $\sqrt{s}=13$ TeV with
    the LHCf Arm2 detector},
   J.\  High Energy Phys.\  {\bf 11}, 073 (2018). 

   \bibitem{ach22} 
S.~Acharya   {\em et al.} (ALICE Collaboration),
  {\em Study of very forward energy and its correlation with particle production at midrapidity 
  in $pp$ and $p$-Pb collisions at the LHC},
   J.\  High Energy Phys.\  {\bf 08}, 086 (2022). 

   \bibitem{ach25} 
S.~Acharya   {\em et al.} (ALICE Collaboration),
  {\em  First observation of strange baryon enhancement with effective energy in pp collisions at the LHC},
   J.\  High Energy Phys.\  {\bf 03}, 029 (2025). 

 \bibitem{ost16a}
S.~Ostapchenko and M.~Bleicher, 
{\em Constraining pion interactions at very high energies 
by cosmic ray data},
Phys.\ Rev.\  D    {\bf 93},   051501(R) (2016).

  \bibitem{aab14}
A.\ Aab et al.\ (Pierre Auger Collaboration), 
{\em  Muons in air showers at the Pierre Auger Observatory:
 Measurement of atmospheric production depth}, 
 Phys.\  Rev.\ D {\bf  90},   012012  (2014).  

\bibitem{kop89} 
 B.\ Z.\ Kopeliovich and B.\ G.\  Zakharov,
{\em Novel mechanisms of baryon number flow over large rapidity gap},
 Z.\ Phys.\  C {\bf 43},  241 (1989).

\bibitem{cap96} 
A.~Capella and B.\ Z.\ Kopeliovich,
{\em Novel mechanism of nucleon stopping in heavy ion collisions},
 Phys.\ Lett.\ B  {\bf 381}, 325 (1996).

\bibitem{kha96} 
D.\ Kharzeev,
{\em Can gluons trace baryon number?},
 Phys.\ Lett.\ B  {\bf 378}, 238 (1996).
 
 \bibitem{dre05}   	
  H.\ J.\ Drescher, A.\  Dumitru,   and M.\  Strikman,  
 {\em High-density QCD and cosmic ray air showers},
     Phys.\ Rev.\ Lett.~{\bf 94},  231801 (2005).	
  
  \bibitem{kai82} 
A.~B.~Kaidalov and K.~A.~Ter-Martirosyan, 
 {\em Pomeron as quark-gluon strings and multiple hadron 
production at SPS collider energies},
 Phys.~Lett.~B   {\bf 117},  247 (1982).

\bibitem{cap94} 
A.~Capella,  U.\ Sukhatme, C.-I.\ Tan, and J.~Tran Thanh Van,
{\em Dual parton model},
 Phys.\ Rep.\   {\bf 236},  225 (1994).
 
 \bibitem{ros77}   	
    G.\ C.\  Rossi   and G.\  Veneziano,
 {\em  A possible description of baryon dynamics in dual and gauge theories},
Nucl.\ Phys.\  B {\bf 123} (1977)   507.
 
\bibitem{col77} 
P.\ D.\ B.\ Collins,
 {\em An introduction to Regge theory and high energy physics},
 Cambridge U.P., 1977.

 \bibitem{aab15}
A.\ Aab et al.\ (Pierre Auger Collaboration), 
{\em Muons in air showers at the Pierre Auger
Observatory: Mean number in highly inclined events}, 
 Phys.\  Rev.\ D {\bf  91},   032003   (2015).  

   \bibitem{aab16}
  A.\ Aab et al.\ (Pierre Auger Collaboration), 
   {\em   Testing hadronic interactions at ultrahigh
  energies with air showers measured by the Pierre Auger Observatory}, 
      Phys.\ Rev.\ Lett.~{\bf 117}, 192001   (2016).	
	
\bibitem{alb22} 
 J. Albrecht, L.\ Cazon, H.\ Dembinski, A.~Fedynitch, and K.-H.\ Kampert,
 {\em  The Muon Puzzle in cosmic-ray induced air showers and its connection
to the Large Hadron Collider}, 
Astrophys.\ Space Sci.\  {\bf 367}, 27 (2022).
 
\bibitem{hil97} 
 A.\ M.\ Hillas,
{\em Shower simulation: Lessons from MOCCA},
        Nucl.\ Phys.\ B Proc.\ Suppl.\   {\bf 52}, 29 (1997).

\bibitem{rei21} 
 M.\ Reininghaus, R.\ Ulrich, and T.\ Pierog,	
{\em Air shower genealogy for muon production},
PoS (ICRC2021),   463 (2021).

 
  \bibitem{ost13}
S.~Ostapchenko,
   {\em QGSJET-II: physics, recent improvements, and results for
  air showers},
  EPJ Web Conf.\   {\bf 52},    02001 (2013).
 
\bibitem{adu17} 
A.\ Aduszkiewicz  {\em et al.} (NA61/SHINE Collaboration),
{\em Measurement of meson resonance production in $\pi^-$-C interactions at SPS energies},
   Eur.\ Phys.\ J.\  C  {\bf 77}, 626 (2017).
    
\bibitem{aga90} 
N.\ M.\ Agababyan  {\em et al.} (EHS-NA22 Collaboration),
{\em Inclusive production of vector mesons in  $\pi^+ p$ interactions 
at 250 GeV/c},
Z.~Phys.~C   {\bf 46}, 387 (1990).
    

\bibitem{aps92} 
    R.\ J.\  Apsimon  {\em et al.} (OMEGA Photon Collaboration),
{\em Comparison of photon and hadron induced production of $\rho^0$ mesons
 in the energy range of 65 GeV to 175 GeV},
Z.~Phys.~C  {\bf 53},  581 (1992).

\bibitem{pie08}
   T.\ Pierog and K.\ Werner,
 {\em Muon production in extended air shower simulations},
        Phys.\  Rev.\  Lett.\    {\bf 101}, 171101 (2008).

\bibitem{man23} 
   J.\ Manshanden, G\"unter Sigl, and  M.~V.~Garzelli,
{\em Modeling strangeness enhancements to resolve the muon excess in cosmic ray 
extensive air shower data},
       JCAP  {\bf 02}, 017 (2023).

\bibitem{adh23} 
H.\ Adhikary  {\em et al.} (NA61/SHINE Collaboration),
{\em Measurement of hadron production in $\pi^-$-C interactions at 158 and 350 GeV/c
 with NA61/SHINE at the CERN SPS},
  Phys.\ Rev.\  D   {\bf 107},  062004 (2023).
  
\bibitem{adh25} 
H.\ Adhikary  {\em et al.} (NA61/SHINE Collaboration),
{\em  Evidence of isospin-symmetry violation in high-energy collisions 
of atomic nuclei},
        Nature Commun.\   {\bf 16},  2849 (2025).

 \bibitem{adu19a} 
A.\ Aduszkiewicz  {\em et al.} (NA61/SHINE Collaboration),
{\em Measurements of hadron production in $\pi^+ + {\rm C}$ and
 $\pi^+ + {\rm Be}$  interactions at 60 GeV/c},
  Phys.~Rev.~D    {\bf 100},  112004 (2019). 

\bibitem{bre82}
A.\ E.\ Brenner  {\em et al.},
{\em  Experimental study of single particle inclusive hadron scattering and associated multiplicities},
  Phys.~Rev.~D    {\bf 26},  1497 (1982). 

\bibitem{bar83}
 D.\ S.\  Barton  {\em et al.},
{\em  Experimental study of the $A$-dependence of inclusive hadron fragmentation},
  Phys.~Rev.~D    {\bf 27},  2580 (1983). 
    
\bibitem{apo10} 
    M.\ Apollonio  {\em et al.} (HARP Collaboration)
{\em Measurements of forward proton production with incident protons and 
charged pions on nuclear targets at the CERN proton synchroton},
  Phys.~Rev.~C    {\bf 82}, 045208 (2010). 
  
\bibitem{agu87} 
 M.\ Aguilar-Benitez  {\em et al.} (LEBC-EHS Collaboration),
{\em Longitudinal distribution of  $\pi^{\pm}$,  $K^{\pm}$, protons 
and anti-protons produced in  360-GeV/c $\pi^-p$ interactions},
 Europhys.\ Lett.\   {\bf 4},  1261 (1987).
   
\bibitem{bau23} 
   S.\ Baur, H.\ Dembinski, M.\ Perlin, T.\ Pierog, R.\  Ulrich, and K.\ Werner,
{\em  Core-corona effect in hadron collisions and muon production in air showers},
  Phys.~Rev.~D    {\bf 107},  094031 (2023). 
\end{thebibliography}
\end{document}